\documentclass[journal=acsnano,manuscript=article]{achemso}

\usepackage[version=3]{mhchem} % Formula subscripts using \ce{}

\newcommand{\rev}[1]{\color{black}{#1}\color{black}}

\usepackage{import}
\usepackage{usepackets/usepackets}
\SectionNumbersOn
\author{Phillip W. K. Jensen}
\affiliation{Chemical Physics Theory Group, Department of Chemistry, University of Toronto, Toronto, Ontario M5G 1Z8, Canada}
\alsoaffiliation{Corresponding author}
\altaffiliation{Contributed equally to this work}
\email{phillip.kastberg@gmail.com}

\author{Lasse Bj\o rn Kristensen}
\affiliation{Chemical Physics Theory Group, Department of Chemistry, University of Toronto, Toronto, Ontario M5G 1Z8, Canada}
\alsoaffiliation{Department of Computer Science, University of Toronto, Toronto, Ontario M5G 1Z8, Canada}
\altaffiliation{Contributed equally to this work}
\email{l.kristensen@mail.utoronto.ca}

\author{Cyrille Lavigne}
\affiliation{Chemical Physics Theory Group, Department of Chemistry, University of Toronto, Toronto, Ontario M5G 1Z8, Canada}
\alsoaffiliation{Department of Computer Science, University of Toronto, Toronto, Ontario M5G 1Z8, Canada}

\author{Al\'an Aspuru-Guzik}
\affiliation{Chemical Physics Theory Group, Department of Chemistry, University of Toronto, Toronto, Ontario M5G 1Z8, Canada}
\alsoaffiliation{Department of Computer Science, University of Toronto, Toronto, Ontario M5G 1Z8, Canada}
\alsoaffiliation{Vector Institute for Artificial Intelligence, Toronto, Ontario M5S 1M1, Canada}
\alsoaffiliation{Canadian Institute for Advanced Research, Toronto, Ontario M5G 1Z8, Canada}
\email{alan@aspuru.com}

\title[An \textsf{achemso} demo]{Towards Quantum Computing with Molecular Electronics}

\newcommand\momentum{p}
\newcommand\lead{\emph{n}}
\newcommand\site{i}
\newcommand\NumberElectrons{\eta}

\keywords{Molecular Computing, Quantum Computing, One- and Two-Electron Scattering Theory, Molecular Electronics, Electronic Structure.}

\begin{document}

\begin{abstract}
In this study, we explore the use of molecules and molecular electronics for quantum computing. We construct one-qubit gates using one-electron scattering in molecules, and two-qubit controlled-phase gates using electron-electron scattering along metallic leads.  Furthermore, we propose a class of circuit implementations, and show initial applications of the framework by illustrating one-qubit gates using the molecular electronic structure of molecular hydrogen as a baseline model. 
\end{abstract}


% Sections -----------------------------------------------------
\section{Introduction}

During the past three decades, the dream of computing using the quantum properties of highly controlled quantum systems has progressed from a theoretical dream to a practical goal pursued by a multitude of companies and academics worldwide. The belief is that this new computational paradigm will be able to solve problems that are intractable on conventional binary computers. Indeed, recent progress has demonstrated the ability of controlled quantum systems to perform tasks beyond the capabilities of even large supercomputers~\cite{arute_quantum_2019,zhong_quantum_2020, wu_strong_2021}. However, many challenges still remain on the road to universal quantum computers capable of running many of the well-known quantum algorithms. Current quantum computing hardware is still limited in size, and tends to be prone to noise that corrupts the stored information if long algorithms are run~\cite{bharti_noisy_2022}. While noise can be managed using quantum error correction~\cite{gottesman_introduction_2009}, doing so will require redundant encoding of the information, thus further reducing the effective size of the algorithms that can be run on the hardware.  For this reason, much work is currently ongoing on how to scale up leading types of quantum hardware~\cite{kottmann_quantum_2021,kyaw_quantum_2021} while dealing with the resulting challenges, such as loss~\cite{takeda_toward_2019, slussarenko_photonic_2019}, crosstalk~\cite{murali_software_2020, rudinger_probing_2019}, input/output bottlenecks~\cite{reilly_challenges_2019, pauka_cryogenic_2019, xue_cmos-based_2021} and the implementation of large-scale cryogenics~\cite{hollister_large_2021, magnard_microwave_2020}. The type and severity of the challenges faced will be different from architecture to architecture. Furthermore, it is possible that the future of quantum computing will feature a mix of different architectures for different purposes. For this reason, new and interesting approaches to quantum computing are still under active investigation~\cite{bartolucci_fusion-based_2021, albert_robust_2020, chiesa_molecular_2020, carretta_perspective_2021}.

In this paper, we explore the use of molecular electronics for quantum computing. This work introduces a working theoretical framework that, admittedly, would require surmunting rather difficult experimental challenges to be realized in a scalable fashion. It is the purpose of this paper to lay out a theoretical framework rather than provide a roadmap to surmount all possible experimental advances needed to realize this vision.

The Hamiltonian considered here is constructed to reflect the division of the system into leads and an interacting electron region, with an additional coupling $\hat{V}$ between the subsystems:

\begin{align}
    \hat{H} = \hat{H}_{\text{leads}} + \hat{H}_{\text{int}} + \hat{V}. \label{eq:system_hamil}
\end{align}
The formalism in \eqref{eq:system_hamil} has been extensively studied in the field of molecular electronics~\cite{meir_landauer_1992}, where the interacting electron region $(\hat{H}_{\text{int}})$ consists of a molecular degree of freedom, typically that of the electronic structure, attached to metallic leads to study, e.g., interference effects~\cite{solomon_understanding_2008, markussen_temperature_2014, andrews_quantum_2008, solomon_quantum_2008, guedon_observation_2012, liu_gating_2017} and heat flow~\cite{segal_thermal_2003, segal_probing_2017} in single molecular junctions. This paper investigates the use of molecular electronics as a basis for quantum computing, using the quantum walk approach pioneered by Farhi \emph{et al.}~\cite{farhi_quantum_1998, farhi_quantum_2008} and Childs \emph{et al.}~\cite{childs_universal_2009, childs_universal_2013}. In their work, a quantum walk refers to the motion of an electron propagating along a chain of states~\cite{childs_example_2002}, and the idea is that interference in the electron motion produces a computation. %Our question is: can molecules produce quantum gates by scattering of electrons?  We address this question by combining the theory of quantum walks and molecular electronics with the goal of realizing a molecular circuit for quantum computing. We believe this work may open a new promising research direction for molecular electronics.

Using molecules in classical electronic circuits was first proposed in 1974 by Aviram and Ratner~\cite{aviram_molecular_1974}. The theory of single-molecule transport, including the role of quantum interference, has developed considerably since~\cite{ratner_brief_2013, xin_concepts_2019, garner_comprehensive_2018, andrews_quantum_2008, borges_effects_2017, segal_electron_2000, segal_thermal_2003, segal_probing_2017}. We are interested in exploring the use of molecules for quantum computing because (1) molecules are extremely small, typically on the order of 1 to 10nm in size, and thus computation with molecules operate within the regime of quantum mechanical effects, (2) Quantum interference effects in molecules are potentially robust to external perturbation~\cite{garner_comprehensive_2018, guedon_observation_2012, gunasekaran_single-electron_2020}, and (3) molecules may be easier and cheaper to fabricate than e.g. superconducting qubits.

Our question is: can scattering of electrons in molecules produce quantum gates? We address this question by combining the theory of quantum walks and molecular electronics with the goal of realizing a molecular circuit for quantum computing. We believe this work may open a new promising research direction for molecular electronics. In this paper, we demonstrate quantum computation using a model system of molecular electron transport under sufficiently cold conditions that vibrations are negligible. Vibrations may turn out to be a hindrance, or their explicit treatment could be exploited to help carry out the vision provided in this work.

The outline of the paper is as follows: In section \ref{sec:one_particle_scattering_theory}, we first explain the type of Hamiltonians used in this work. We then propose a one-particle scattering theory, and outline its connection to one-qubit gates. In section \ref{sec:two_particle_scattering_theory} we propose a two-particle scattering theory for the implementation of the controlled-phase gate. Next, in section \ref{sec:molecular_circuits} we show an example molecular circuit, specifically a Hadamard test circuit. Finally, conclusion and outlook can be found in section \ref{sec:conclusion_and_outlook}.

\section{One-particle scattering theory}
\label{sec:one_particle_scattering_theory}

Guided by the typical experimental geometry of molecular electronics setups, we view the incoming electrons in the leads as moving freely, since if the leads are considered to be metallic, then the interaction between electrons in the leads is strongly screened and can be neglected. In the case of continuous space, the free-particle Hamiltonian would be

\begin{align}
 \hat{H}_{\text{free}} = \frac{\hat{p}^2}{2m}, \label{eq:con_hamil}
\end{align}
where $\hat{p}$ is the momentum operator and \emph{m} is the mass of the electron. However, the system Hamiltonian \eqref{eq:system_hamil} consists of leads connected to a molecular Hamiltonian. Since the solution of the electronic Schr\"{o}dinger equation often begins with the discretization of the wave function, the molecular Hamiltonian will be assumed discretized, and it seems conceptually cumbersome to combine a continuous Hamiltonian with a discretized one. Therefore, to overcome this, we propose a discretized free-particle Hamiltonian on the form

\begin{align}
 \hat{H}_{\text{leads}} = \sum^N_{\lead =1} \sum^\infty_{\site = 1} \bigg(  2 \beta \hat{c}^\dagger_{\lead,\site} \hat{c}_{\lead,\site}  -\beta \big(\hat{c}^\dagger_{\lead,\site} \hat{c}_{\lead,\site + 1} + \text{h.c.} \big) \bigg), \label{eq:dis_hamil}
\end{align}
where $\hat{c}^\dagger_{\lead,\site}$ ($\hat{c}_{\lead,\site}$) creates (destroys) an electron in the localized site $\site$ along the lead $\lead$. The model is often used to describe an idealized physical system of an infinite line consisting of discrete quantum dots with a single localized orbital at each dot, and an electron moving along the line. In that case, the hopping energy $\beta$ is a fixed, system-dependent coupling strength between the dots. In our case, however, we see \eqref{eq:dis_hamil} as an abstract parametrized model that can be mapped to a free-particle model, and $\beta$ as some free parameter. The Hamiltonian \eqref{eq:dis_hamil} is particularly useful because it closely resembles that of a free-particle model due to sharing the same eigenstates (up to a discretization), but the eigenenergies are not the same. This is expected since after all they are two different models. However, the two models  \eqref{eq:con_hamil} and \eqref{eq:dis_hamil} are identical (up to a discretization) for all $\abs{\momentum} < \abs{\momentum_0}$ if we choose the parameters such that $a \ll   \hbar / \abs{\momentum_0}$ and $\beta = \hbar^2 / 2ma^2$, where \emph{a} is the distance between the dots, and $\momentum_0$ is some eigenvalue of the discretized momentum operator (supporting information \ref{app:kinetics}). That is, we require that the distance between the dots of the model is much smaller than the inverse of the incoming electron momentum in order to be able to map between the free-particle model and the discretized model. \rev{Note, the distance between the dots `\emph{a}' is not a physical distance  but a parameter of an abstract model. For example, if  $ p_0 = \hbar/a_0$, i.e., corresponding to $p^2_0/2m_e = 13.6$ eV  --- the energy scale of the incoming electrons --- then the parameter must be much smaller than the Bohr's radius, i.e., $a \ll a_0$, such that we can interpret $p_0$ as the incoming momentum of a free particle. }

The interacting electron region is described by the molecular non-relativistic electronic Hamiltonian within the Born–Oppenheimer approximation:

\begin{align}
\hat{H}_{\text{int}} = \sum^M_{p,q=0}   h_{pq} \hat{a}_{p}^\dagger \hat{a}_{q} + \frac{1}{2}  \sum^M_{p,q,r,s=0}  h_{pqrs}  \hat{a}_{p}^\dagger  \hat{a}_{r}^\dagger \hat{a}_{s} \hat{a}_{q} + h_{\text{nuc}}, \label{eq:int_hamil}  
\end{align}
where $\{\hat{a}_p\}$ and $\{\hat{a}^\dagger_p\}$ form a set of single-electron creation and annihilation operators in the molecule, \emph{M} is the number of spin-orbitals, and $h_{pq}$ and $h_{pqrs}$ are one- and two-electron integrals in Dirac notation—the one-electron integrals involving the electronic kinetic energy and the electron-nuclear attraction, and the two-electron integrals involving the electron–electron interaction (see for instance Ref. \cite{helgaker_molecular_2000}). The scalar term, $h_{\text{nuc}}$, represents the nuclear-repulsion energy. Note that while the leads are described using a localized orbital basis representation, the molecule may be in a delocalized one, e.g., consisting of the delocalized molecular orbitals. 

Finally, the coupling that allows the actual transfer of charge between the subsystems is described by

\begin{figure}[t] 
\centering  
\includegraphics[width=1.0\textwidth]{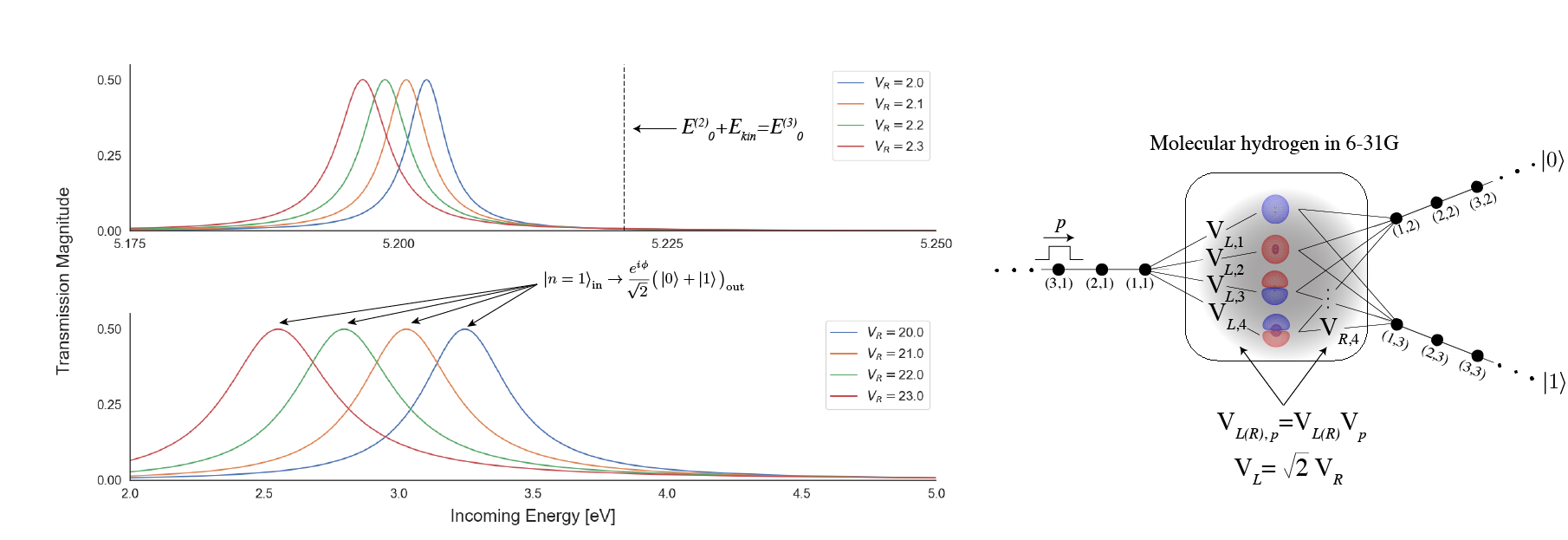}
\caption{\rev{Electron transmission magnitude $\abs{T_{L,R}(\momentum)}^2$ as a function of incoming kinetic energy $E_{\text{in}}= \frac{\momentum^2}{2m}$ for molecular hydrogen in the 6-31G basis attached between three leads; one input and two output leads, for various lead-molecule coupling strengths. The illustration shows the setup, where the molecular orbitals shown are the four orbitals from $\text{H}_2$ in the 6-31G basis, which are attached to the leads with coupling strength $V_{\lead,p}$. The quantum dots along the leads are labeled as $(\lead,\site)$, where $\lead$ and $\site$ indicate the lead and site, respectively, and we set the quantum dot distance to $a = 0.1\text{\AA}$.  }}
\label{fig:introduction_illustration}
\end{figure}

\begin{align}
\hat{V} &= \sum^N_{\lead = 1} \sum^M_{p=0}  \big( V_{\lead, p} \hat{c}^\dagger_{\lead,1} \hat{\tilde{a}}_{p} + V_{\lead, p}^* \hat{\tilde{a}}^\dagger_{p} \hat{c}_{\lead,1} \big), \label{eq:V_hamil} \end{align}
where $\hat{\tilde{a}}_p = \hat{s}^\dagger \hat{a}_p $, and $\hat{s}$ is an operator that generate molecular orbitals with the lowest energy for the charged molecule from those of the uncharged molecule, see supporting information \ref{app:lead_molecule} for details. To evaluate the coupling strengths $V_{\lead, p}$ numerically, we assume that $V_{\lead, p} = V_{\lead} V_{p}$ and that $V_{p}$ equals the absolute square of the molecular orbital which takes part in the transfer process, integrated over a region close to the leads (supporting information \ref{app:lead_molecule}). It seems reasonable that if the molecular orbital density is very small near the lead $\lead$, then consequently the coupling strength through this particular orbital should be small as well. \rev{That is, low density should imply weak coupling. Similarly, if the molecular orbital density is large near the lead $\lead$, then the coupling strength should be large}. As a remark, the theory proposed here is not limited to our description of the coupling strengths, and can generalize to more sophisticated approaches to calculating the coupling strengths.

To calculate charge transfer, we use standard scattering theory. Following Refs. \cite{childs_levinsons_2011,childs_levinsons_2012, childs_universal_2009, childs_universal_2013} for quantum computing by quantum walks, we propose the one-particle scattering ansatz:

\begin{align}
&\ket{\text{sc}_{\lead}(\momentum)} = \sum^N_{\lead' = 1} \sum^{\infty}_{x = 1} \big(e^{-i \frac{\momentum a x }{\hbar}} \delta_{\lead,\lead'} + e^{i \frac{\momentum a x }{\hbar}} S_{\lead,\lead'}(\momentum)  \big) \ket{\lead', x; E^{(\NumberElectrons)}_0} + \sum^G_{g=0} w_{\lead,g}( \momentum ) \ket{E^{(\NumberElectrons+1)}_g},  \label{eq:one_particle_sc}
\end{align}
where we have denoted the electron with momentum $\momentum$ being at position \emph{x} along chain $\lead'$ and the molecule being in its neutral charged ground state (with $\NumberElectrons$ electrons) by the ket $\ket{\lead', x; E^{(\NumberElectrons)}_0}$, and the electron occupying the energy eigenstate \emph{g} of the $-1$ charged molecule by the ket $\ket{E^{(\NumberElectrons+1)}_g}$. Along the leads, the factors  $\exp( \pm i \momentum a x / \hbar) $ are picked up by the wavefunction in \eqref{eq:one_particle_sc} when applying the discretized translation operator to move the electron. Additionally the sign in the exponent determines the direction of travel of the moving electron. In this interpretation, the object $S_{\lead,\lead'}(\momentum)$ corresponds to the amplitude with which the electron is either transmitted along a new lead $(T_{\lead,\lead'}(\momentum) = S_{\lead,\lead' \neq \lead}(\momentum))$ or reflected back along the lead it came from $(R_{\lead}(\momentum) = S_{\lead,\lead' = \lead}(\momentum))$. Notice that the scattering states have a well-defined  momentum  at the cost of an ill-defined position. We require that these states fulfill the eigenvalue equation  $\hat{H} \ket{\text{sc}_\lead(\momentum)} = (\momentum^2/2m + E^{(\NumberElectrons)}_0) \ket{\text{sc}_\lead(\momentum)}$, where $\momentum^2/2m$ and $E^{(\NumberElectrons)}_0$ are the incoming kinetic energy and initial molecular ground state energy, respectively. This equation can be solved for the set of coefficients $\{S_{\lead,\lead'}\}$ and $\{w_{\lead,g}\}$, see supporting information \ref{app:One_Particle_Scattering_Matrix} for details. Thus for a given system Hamiltonian, both the transmission and reflection amplitudes can be calculated.

An example of a computed set of scattering coefficients is given in Fig. \ref{fig:introduction_illustration}, where we show electron transmission magnitude as a function of incoming kinetic energy for molecular hydrogen in the 6-31G basis attached between three leads; one input and two output leads, for various lead-molecule coupling strengths. To evaluate the coupling strengths numerically, we let $V_{\lead,p} = V_{\lead}V_{p}$ and use Eq. \eqref{app:eq_V_int} to evaluate $V_{p}$ choosing to integrate up to $0.1a_0$ ($a_0$: Bohr's radius) from the protons. For clarification, see Fig. \ref{fig:H2_STO3G_transmission}, where the plane intersections indicate the integration limits. To achieve high transmission, we set $V_{L} = \sqrt{2} V_{R}$ (supporting information \ref{app:example_independent_from_leads}), where $V_{L} \equiv V_{\lead = 1}$ and  $V_{R} \equiv V_{\lead = 2} =V_{\lead = 3}$, and compute the transmission magnitude $\abs{T_{L,R}(\momentum)}^2$ for various values of $V_R$. Since the output coupling strengths are identical, the transmission amplitude for each output lead is identical, $  T_{1,2}(\momentum) =T_{1,3} (\momentum)$. Molecular hydrogen may not be a realistic system to attach between leads due its very small size, but it is likely that many molecules attached to three leads would be able to create an equal superposition of the outgoing electron, see supporting information \ref{app:example_independent_from_leads} for details. We observe that at certain incoming energies, the transmission magnitude equals 1/2 in both output leads, and we can write the output state as $\ket{\lead = 1}_{\text{in}} \rightarrow \frac{e^{i\theta}}{\sqrt{2}} \big(\ket{\lead = 2} +\ket{\lead = 3} \big)_{\text{out}}$, which is an equal superposition of the two output leads traversed by the outgoing electron, up to some global phase. The global phase is given by the argument of the transmission amplitude, $\phi = \arg (T_{L,R})$. Note that, the transmission peaks are not located at $E^{(3)}_0 - E^{(2)}_0 $ (vertical dashed line in Fig. \ref{fig:introduction_illustration}), where $E^{(2)}_0$ and $E^{(3)}_0$  are the ground state energies for the neutral and charged molecule, respectively, as a consequence of the leads ``mixing'' with the energy eigenstates of the molecule and, consequently, shifting the transmission channels. This may come as a surprise since at an incoming energy of $E^{(3)}_0 - E^{(2)}_0 $, it matches that of the ground state energy of the charged molecule, and we would expect high transmission. We refer the readers to supporting information \ref{app:example_h2_transmission} for a more detailed explanation of how to compute the electron transmission. 

In the context of quantum computing, we define the state of the qubit by the position of the electron by denoting the state as $\ket{0}$ if the electron is on the upper lead, and $\ket{1}$ for the lower lead, as depicted in Fig. \ref{fig:introduction_illustration}~(A). For zero reflection, the setup depicted in  Fig. \ref{fig:introduction_illustration}~(A)  can be used to initialize qubits in an equal superposition state (up to a global phase), which is a useful building block for many quantum algorithms, e.g., the quantum phase estimation~\cite{abrams_simulation_1997, abrams_quantum_1999, kitaev_quantum_1997} and Grover's search algorithm~\cite{grover_quantum_1997, grover_fast_1996}. In this work, we will use this setup for the Hadamard test circuit~\cite{aharonov_polynomial_2009, mitarai_methodology_2019} in section \ref{sec:molecular_circuits}.

\section{Two-particle scattering theory}
\label{sec:two_particle_scattering_theory}

\begin{figure}[t] 
\centering  
\includegraphics[width=1.0\textwidth]{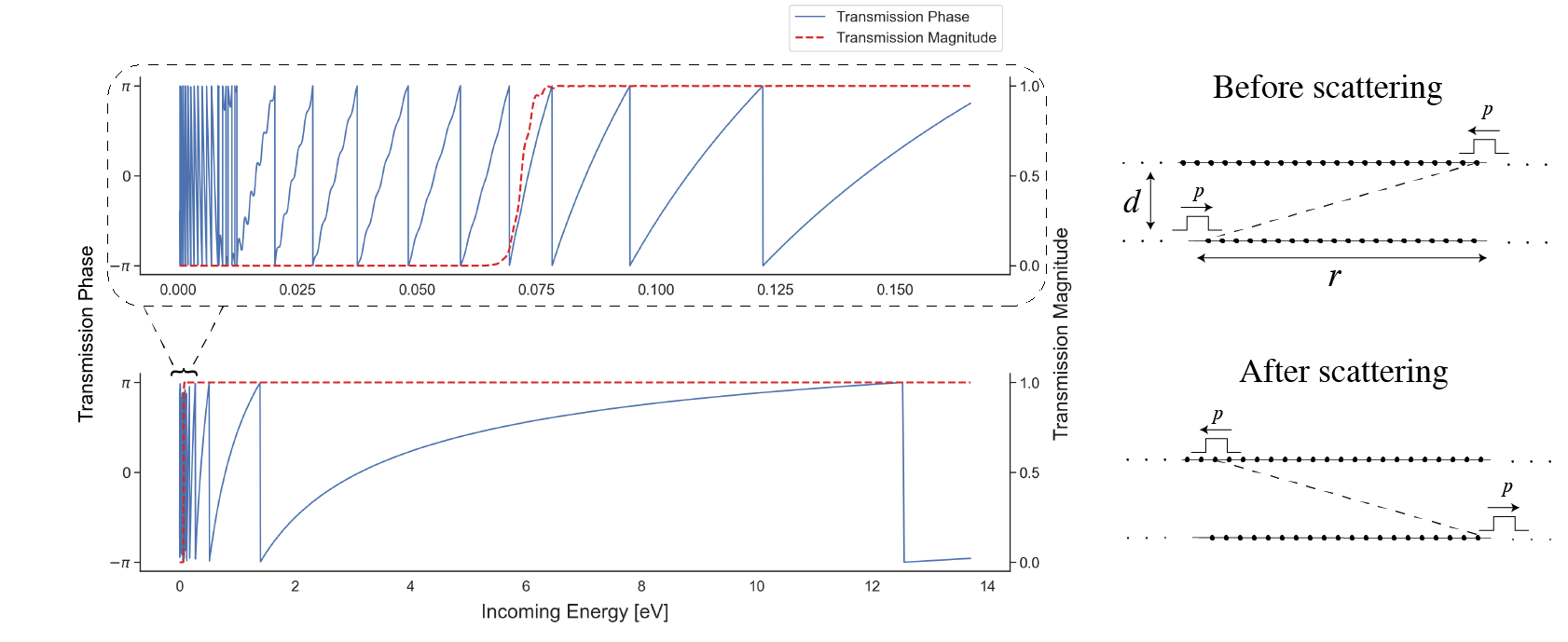}
\caption{\rev{Two-electron transmission magnitude and transmission phase $\arg(T(\momentum))$ as a function of incoming kinetic energy when the electrons move past each other on separate leads while interacting, as depicted in the illustration. We set $a = 0.1\text{\AA}$, $d = 1000$ and the cut off to $C = 10000$. } }
\label{fig:two_particle_scattering}
\end{figure}

The one-particle scattering theory is sufficient to construct one-qubit gates, but to really access the power of quantum computation we need to consider electron-electron interactions. In order to implement two-qubit entangling gates, we route two electrons close to each other, let them interact, and separate them again, as depicted in the illustration in Fig. \ref{fig:two_particle_scattering}. A similar strategy was first proposed in Ref. \cite{childs_universal_2013}, where they considered scattering of two electrons on the same lead. Our starting point is a Hamiltonian with the following contributions

\begin{align}
\hat{H} &=\hat{H}_{\text{leads}}  +  \hat{H}_{\text{int-leads}}, \label{eq:hamil_twobody_scatt}
\end{align}
where $\hat{H}_{\text{leads}}$ is the one-body Hamiltonian \eqref{eq:dis_hamil}, and the two body interaction is given by the Coulomb repulsion term,

\begin{align}
 \hat{H}_{\text{int-leads}} = \sum_{ \site_1,\site_2 \in \mathbb{Z}}V(r,d) c_{\lead_1, \site_1 }^\dagger c_{\lead_2, \site_2 }^\dagger c_{\lead_2, \site_2 } c_{\lead_1, \site_1 }, \label{eq:two_body}
\end{align}
with $\lead_1 \neq n_2$ as the lead indices, and

\begin{align}
V(r,d) =
    \begin{cases}
       K \frac{1}{\sqrt{r^2 + d^2}} & \text{if $C\geq \abs{r}$}\\
      0 & \text{if $C< \abs{r}$}.
    \end{cases}      \label{eq:potential}
\end{align}
Here $r = \site_1 - \site_2 $ is the horizontal distance between the electrons, $d\in \mathbb{Z}$ is the distance between the two leads $\lead_1$ and $\lead_2$ in units of \emph{a}, and $K = kq^2/a$ ($k=\text{Coulomb constant}$, $q=-e$, $a=\text{quantum dot distance}$). The two-body term \eqref{eq:two_body} assumes the electrons cannot jump to other leads, but that they can still interact while moving past each other on separate leads. The justification for this form is that the electrons would have to overcome a very large potential barrier in order to jump to other leads. Furthermore, we assume the potential \eqref{eq:potential} has finite range \emph{C}, which means that $V(r,d) = 0$ whenever $C< \abs{r}$ (same assumption as in Ref. \cite{childs_universal_2013} in SI section S2). We refer the reader to supporting information \ref{app:S_matrix_two_particles} for a more detailed discussion. 

Figure \ref{fig:two_particle_scattering} shows two-electron transmission magnitude and phase when letting the electrons past each other on separate leads while interacting. The electrons are assumed to have the same kinetic energy, and for the potential \eqref{eq:potential} we set $C = 10^4$ and $d = 1000$ in units of \emph{a}, which result in a separation of $1000a = 10$ nm. For this lead-lead distance of 10 nm, we observe that for a few millielectronvolts the electrons have overcome the potential repulsion and they are fully transmitted. The transmission phase, which is the argument of the transmission amplitude, oscillates as a function of kinetic energy, and by letting the electrons simply pass each other with kinetic energy greater than a few  millielectronvolts, we obtain a fully transmitted output state but with a phase shift, $\ket{\lead_1,\lead_2}_{\text{in}} \rightarrow e^{i\phi} \ket{\lead_1,\lead_2}_{\text{out}}$, where $\phi = \text{arg} (T)$. The transmission phase oscillates fast at low energies because the electrons are spending more time near each other and interact more. Furthermore, if the distance between the leads gets very large, then the electrons are essentially non-interacting $V(r,d) \approx 0$, the transmission magnitude becomes a step function, and the phase goes to zero, as can be seen from Fig.  \ref{fig:app:two_particles_scattering_numerics} (supporting information \ref{app:S_matrix_two_particles}). We will use this setup to construct the two-qubit controlled-phase gate by simply letting electrons pass each other on separate leads. 

\section{Molecular circuits for Quantum Computing}
\label{sec:molecular_circuits}

We pick an encoding where the state of the qubit is represented by the position of the electron. Each qubit consists of two leads, and we denote it as the state $\ket{0}$ if the electron is on the upper lead, and $\ket{1}$ on the lower lead,  as depicted in Fig. \ref{fig:introduction_illustration}. We can then construct one-qubit gates by adding molecules between the leads of the qubit, and two-qubit gates by routing two electrons closer to each other, letting them interact, and separating them again. 

In the following, we are going analyze the molecular circuit depicted in Fig. \ref{fig:molecular_circuit}, which corresponds to the Hadamard test circuit~\cite{aharonov_polynomial_2009, mitarai_methodology_2019}. The first step is to initialize the first qubit in the state $\ket{+} = \frac{1}{\sqrt{2}}(\ket{0} + \ket{1})$. For this, we require a molecule attached to three leads, one for input and two for output. Furthermore, the transmission magnitude should equal 1/2 for each output lead, hence we need zero reflection. This was observed for molecular hydrogen in Fig. \ref{fig:introduction_illustration} at certain energies. The theory presented here is not limited to molecular hydrogen, and likely a large class of molecules attached to three leads may work. In the case of having four leads attached to the molecule, we show in supporting information \ref{app:example_independent_from_leads} that for any real couplings fulfilling our simplifying assumptions, there will always be some reflection, and thus it seems not to be possible to make an equal superposition state. Moving along the circuit, we note that after scattering of the first molecule, the system state  is $\frac{e^{i\theta} }{\sqrt{2}} ( \ket{0} + \ket{1} )\otimes \ket{q} $, as depicted in Fig. \ref{fig:molecular_circuit}. The global phase comes from the scattering of the molecule and is given by $\theta = \arg(T_{L,R})$, where $T_{L,R}$ is the transmission amplitude through the molecule. Due to the  electrons interacting, a relative phase is then introduced to the system state. The  electron-electron interaction occurs only if the electrons are close to each other, and in qubit notation the resulting operation reads: 

\begin{figure}[t] 
\centering  
\includegraphics[width=1.0\textwidth]{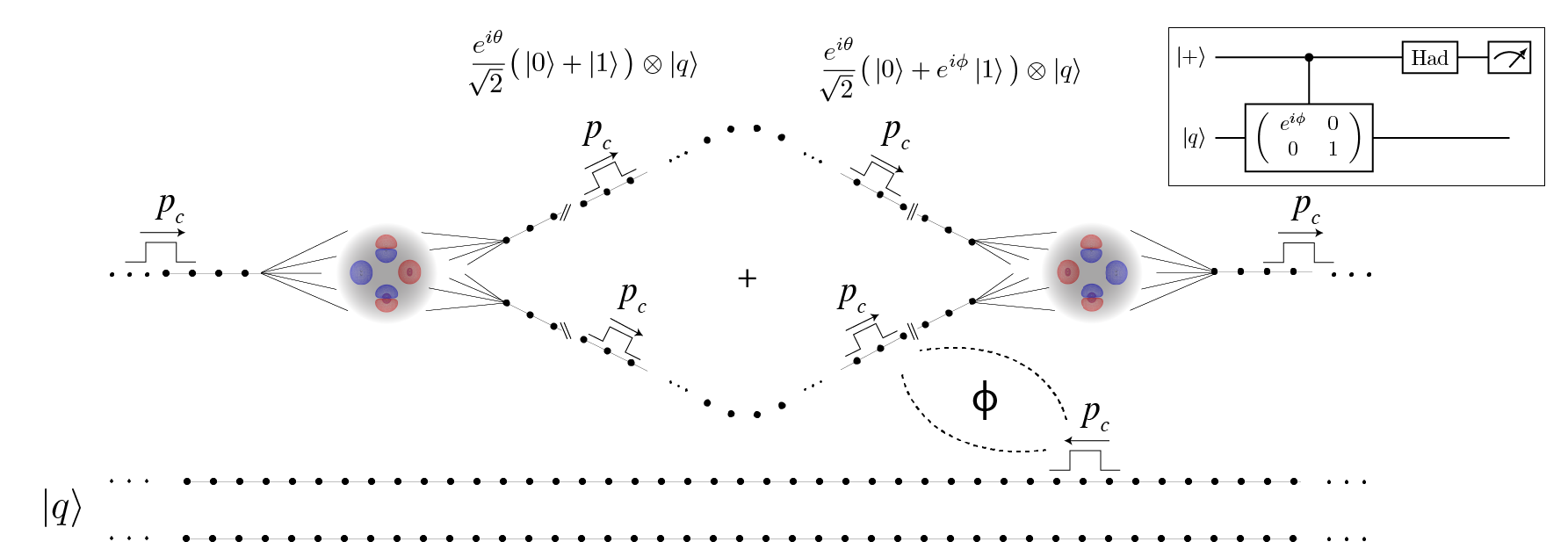}
\caption{ The molecular (Hadamard test) circuit. The insert quantum circuit illustrates the Hadamard test circuit, specifically that for the unitary $\text{diag}(e^{i\phi},1)$, where ``Had'' indicates the Hadamard gate. We denote $\momentum_c$ as the computational momentum, assuming all other momenta are filtered out before entering the circuit. The phase $\phi$ is given by the argument of the transmission amplitude from the two-electron scattering process. The (+) symbol indicates the electron is in a superposition of being on the upper and lower leads.  }
\label{fig:molecular_circuit}
\end{figure}

\begin{equation}
\begin{rcases}
 \ket{10}_{\text{in}} \rightarrow e^{i\phi} \ket{10}_{\text{out}} \\
\ket{00}_{\text{in}} \rightarrow  \ket{00}_{\text{out}} \\
\ket{01}_{\text{in}} \rightarrow   \ket{01}_{\text{out}}  \\
\ket{11}_{\text{in}} \rightarrow   \ket{11}_{\text{out}}
  \end{rcases}
  \text{CPHASE} \label{eq:CHPASE}
\end{equation}
where CPHASE stands for the controlled-phase gate. According to  Eq. \eqref{eq:CHPASE}, a phase is multiplied to the state $\ket{10}_{\text{out}}$ due to the electron-electron interaction, and we show the transmission phase $(\phi)$ as a function of incoming kinetic energy in Fig. \ref{fig:introduction_illustration}~(B). Thus immediately before scattering on the second molecule, the system state reads $\frac{e^{i\theta} }{\sqrt{2}} ( \ket{0} + e^{i\phi} \ket{1} )\otimes \ket{q} $ if $q = 0$, as depicted in Fig. \ref{fig:molecular_circuit}. The final transmission magnitude for finding the electron on the rightmost lead is then given as

\begin{align}
\abs{T_f}^2 = \frac{1}{2} \big( 1 + \cos( \phi) \big),\label{eq:final_transmission}
\end{align}
as we derive in supporting information \ref{app:sec:transmission_probability_of_molecular_circuit}. This is an interesting result since it implies that the output reflects non-trivial superposition and interaction effects. We can change the current by varying the distance between the interacting electrons, which will then change $\phi$. Similarly, the phase will depend on the state of the bottom qubit in Fig. \ref{fig:molecular_circuit}, being approximately zero if the bottom qubit is in the $\ket{1}$ state. In this case $ \phi = 0$ (i.e. the electrons are independent of each other), the transmission magnitude equals one, as expected. We notice that the transmission magnitude \eqref{eq:final_transmission} is equivalent to the probability that a measurement yields the zero state for the quantum circuit depicted in Fig. \ref{fig:molecular_circuit}, which is the Hadamard test circuit~\cite{aharonov_polynomial_2009, mitarai_methodology_2019}. Thus, effectively our molecular circuit does quantum computation, specifically a Hadamard test. Indeed, two-qubit gates of the form \eqref{eq:CHPASE} are universal for quantum computation when paired with arbitrary single-qubit gates~\cite{bremner_practical_2002}.

\rev{The molecular circuit in Fig. \ref{fig:molecular_circuit} was just one example of a molecular circuit performing quantum computing. For example, it should be possible to construct the Pauli gate set $\{X,Y,Z\}$ using our setup. To construct the \emph{X}-gate, we simply route the electron from the upper to the lower lead, $\ket{0}\rightarrow \ket{1}$, or lower to upper, $\ket{1} \rightarrow \ket{0}$, without the need of any molecule. For the \emph{Z}-gate, we would need to incorporate a molecule on the lower lead such that the electron is fully transmitted but acquire a phase of $\arg(T) = e^{i\pi}$. Finally, the \emph{Y}-gate can then be realized (up to a global phase) as the product $XZ = -i Y$. Further, the \emph{T}-gate can be constructed in a similar way as the \emph{Z}-gate. To construct a universal gate set, a Hadamard gate is required. As pointed out earlier, the Hadamard gate seems to be difficult to construct due to the finite reflection coefficient in the output lead  (supporting information \ref{app:example_independent_from_leads}). We got around this problem in Fig. \ref{fig:molecular_circuit} using a `combined' Hadamard gate and measurement, which works for this particular circuit.  We believe that finding molecular candidates for this setup is a very important question. Thus, finding potential molecular candidates for quantum computing  is an open question for future work.

While not necessary for universality, tunable gates would be useful for, e.g., quantum variational algorithms. To construct tunable gates we would need to change the electronic structure of the molecules. It seems impractical to physically exchange the molecules in the circuit, but affecting the electronic structure of the molecules while remaining attached to leads may work, e.g., applying an external magnetic field. We leave the question of tunable gates for further work.  }  

\section{Conclusion and Outlook}
\label{sec:conclusion_and_outlook}

In this work, we investigated the use of molecules and molecular electronics for quantum computing. Our proposal was based on the papers on quantum computing by quantum walks~\cite{childs_universal_2009, childs_universal_2013}, where electron scattering produces the computation. We developed a theory to encode one-qubit gates into scattering of electrons in molecular junctions and two-qubit controlled-phase gates into scattering of electrons on separate leads. We recognize there is still a long way ahead before achieving a functional quantum chemical computer, but we believe this work may be one of the initial steps towards the formalization of quantum computing with molecular electronics. In the appendices below, we discuss limitations and further directions for constructing molecular gates.

\rev{ Firstly, we want to acknowledge related work using molecular physics for quantum computing. In our earlier work, we proposed molecular systems to realize the classical  \textsc{NAND} gate, the universal gate for classical digital computing, using the same principles as this work, i.e., one-electron scattering theory~\cite{jensen_molecular_2018}. Other related setups was proposed in refs.~\cite{albert_robust_2020,chiesa_molecular_2020} by encoding the qubit in the rotational states of a diatomic molecule~\cite{albert_robust_2020} or energy levels of molecular nanomagnets~\cite{chiesa_molecular_2020}.      }

\rev{One potentially interesting connection is the relation between the scheme presented here and the concepts used in optics-based quantum computing. In both cases, the information carriers constituting the qubits are travelling degrees of freedom, and gates are implemented by sequences of objects that they interact with throughout their path. A central difference is that using charged particles instead of the uncharged bosonic photons should allow for easier implementation of interactions and 2-qubit gates, though potentially at the cost of more sensitivity to external noise. Nevertheless, many ideas developed in an optical context are likely to be useful here, for instance quantum teleportation-based protocols for mitigating the non-deterministicness of operations~\cite{knill_scheme_2001,grice_arbitrarily_2011}. Of special interest is the concept of measurement-based quantum computation, which allows for universal quantum computation using only fixed resource-state generation and adaptive readout~\cite{raussendorf_one_2001, briegel_measurement_2009, bourassa_blueprint_2021,bartolucci_fusion-based_2021}. In our context, this corresponds to achieving universality while requiring minimal changes to the molecular circuit, since the entire state-preparation pipeline would be fixed. In fact, the central building blocks for this resource-state generation --- $\left| + \right>$-state generation and CZ gates --- seems to be native building blocks of the proposed architecture. However, implementations of the required adaptive measurements is still an open question that needs to be solved by further research before universality can be achieved.}

Throughout this work, we made some assumptions for our description of the molecular junction, and we summarize them in supporting information \ref{sec:app:List_of_Approximations}. Perhaps the most profound of them is to not include molecular vibrations. Transitions between vibrational modes happen at room temperature, meaning at this level of theory we would require cooling of the molecules to reduce the effect of molecular vibrations. A detailed description of the effects of vibrations will be included in an upcoming paper. Furthermore, we recognize that the idea of attaching a molecule between metallic leads may be highly nontrivial to realize in practice, hence the idea of doing quantum computation by single molecular junctions may not be feasible for experimental realization. There may be other related ways of realizing this idea, as discussed below, and our hope is that this work will open a discussion of potential experimental setups to realize molecular gates for quantum computing.

It was assumed that the initial electronic molecular state before scattering is the ground state $\ket{E_0^{(\NumberElectrons)}}$ (see Eq.\eqref{eq:one_particle_sc}). We know that the probability of being in the electronic ground state is $P_0 = 1/Z$  ($E_0^{(\NumberElectrons)} \equiv 0$), where \emph{Z} is the partition function, $Z = \sum^G_{g = 0}  e^{-\beta E_g^{(\NumberElectrons)}}$. Similarly, the probability of being in the first excited state is $P_1 = e^{-\beta E_1^{(\NumberElectrons)}}/ Z$. For the molecule to be found in the electronic ground state with high probability, we require that $1 \gg e^{-\beta E_1^{(\NumberElectrons)}}$, or equivalently $1 \ll \beta E_1^{(\NumberElectrons)}$. The thermal energy at room temperature is $k_B T = 0.025$ eV, hence $\beta = 40$ eV$^{-1}$, and the energy gap between the electronic ground and first excited state is typically in the electronvolt range, so indeed the molecule will most likely be in the electronic ground state at room temperature. As pointed out earlier, we did, however, not include molecular vibrations in our descriptions, and transitions between vibrational modes happen in the meV range. At room temperature, the molecule will therefore vibrate due to the thermal energy, and this affects the transmission through the molecule. For this reason, cooling of the molecules may be necessary. We leave a detailed analysis of these effects for further work.

What type of noise is our molecular device most prone to experience? Amplitude damping is the process where a qubit decays to its lower state, $\ket{1} \rightarrow \ket{0}$. It seems unlikely that an electron would jump to another lead, i.e., bit flip errors seem unlikely, but other processes may cause similar effects. For example, the electron could be trapped in the lead or within the molecule, making the qubit disappear, $\ket{1} \rightarrow \text{no qubit}$ or $\ket{0} \rightarrow \text{no qubit}$. This type of noise should be easily detectable, and hence easy to correct for using postselection. More problematic is phase errors. For example, impurities along the leads may cause transmission phases to be added in an incoherent way by essentially adding extra artificial dots along the lead, causing the phase to be shifted. We leave the question of noise and its effect on molecular gates as future work.

What is the effect of the metallic leads on the electronic structure of the molecule beside shifting the channels? In  Ref. \cite{olivares-amaya_anion_2011}, the authors modeled a benzene molecule between two metallic plates, and calculated the total free energy of the system for different voltage bias and orientations of the molecule theoretically. Through these studies, the authors observed several nontrivial effects. For instance, orientational effects are important for the stability of the molecule in the junction, and thus orientational effects may cause significant change in the transmission spectrum. Thus, more detailed description of the interaction between the leads and the molecule may be required. We leave this question as future work.

One thing to note is that the metallic leads did not play an essential role in the computation, and could in principle be replaced by other systems capable of charge conduction. For example, attaching a molecule between metallic leads may be a highly nontrivial to realize in practice, and inherent energy-mismatches could potentially hinder transmission. One promising alternative candidate would be strands of DNA, due to their relative ease of synthesis and ability to bond in a controlled way to both molecules and metallic electrodes~\cite{arnold_dna_2016,slinker_dna_2011}. While the level of coherence involved in the transport is still an open question, and conductivity can be sensitive to perturbations~\cite{arnold_dna_2016}, recent experiments demonstrating long-range charge transport~\cite{tse_effective_2019} nevertheless hint at the interesting possibility of a fully molecular architecture built from molecules interconnected by DNA wires.

Lastly, we recognize that the state-of-the-art procedure to calculate the transfer of charge in molecular junctions is to compute it as $J=\braket{\dot{\hat{N}}}$, where \emph{J} is the current and $\hat{N}$ is the number operator for the leads. In Ref. \cite{meir_landauer_1992}, it was shown that the equation for the current can be written very neatly in terms of Green's functions for the interacting electron region and the leads (Eq. (8) in \cite{meir_landauer_1992}). Furthermore, they described the leads using the Fermi–Dirac distribution, so that what was left to solve in order to calculate the current through molecular junctions was "simply" the Green's functions for the isolated molecule. However, it seems there is no clear connection between the transmission amplitude and the formalism in Ref. \cite{meir_landauer_1992}, unless the molecular Hamiltonian is non-interacting, where a connection does arise. A non-interacting Hamiltonian is not an option here if we want to design realistic molecular gates. That being said, there may be a way to use the formalism in Ref. \cite{meir_landauer_1992} to compute the transmission and transmission phase in order to construct molecular gates, but we leave this question for further work. 

\section{Acknowledgement}

We thank Philipp Schleich, Thi Ha Kyaw and Robert Pollice for detailed feedback on the manuscript. Al\'an Aspuru-Guzik and his research group acknowledge the generous support from Google, Inc. in the form of a Google Focused Award. A.A.-G. also acknowledges support from the Canada Industrial Research Chairs Program and the Canada 150 Research Chairs Program, the Vannevar Bush Faculty Fellowship under contract ONR N00014-16-1-2008, and the Natural Sciences and Engineering Research Council of Canada (NSERC). P.W.K.J acknowledges support by Augustinus Fonden, Knud H\o{}jgaards Fonden, and Viet-Jacobsens Fonden. The Quantikz package was used for typesetting the quantum circuit in Fig. \ref{fig:molecular_circuit} with \LaTeX~\cite{kay_quantikz_2019}. 
\begin{suppinfo}

Comparison of kinetic term in discretized and continuous 1D models, an analysis of the lead-molecule coupling, the one- and two-particle scattering matrix, antisymmetrization of the two-particle scattering state, derivation of the final transmission magnitude of the molecular circuit, and table of assumptions. 

\end{suppinfo}

% % % Appendix -----------------------------------------------------
\counterwithin{figure}{section}
\section*{Appendices}
\appendix
\section{Comparison of Kinetic Term in Discretized and Continuous 1D Models}
\label{app:kinetics}

Often, discretized models of an electron moving along a semi-infinite chain in one dimension is represented using a Hamiltonian of the form 
\begin{align}
\hat{H}_{\text{dis}} &= \beta\sum_{\site \in \mathbb{Z}} \big( \hat{c}^\dagger_{a\site} \hat{c}_{a\site+a} +\hat{c}^\dagger_{a\site+a} \hat{c}_{a\site} \big), \label{app:eq:H_dis}
\end{align}
where \emph{a} is the distance between the discretized possible positions of the electron, e.g., the quantum dots in the chain that it is moving through. We can write the Hamiltonian \eqref{app:eq:H_dis} in terms of the discretized translation operators

\begin{align}
\hat{H}_{\text{dis}} &=  \beta\big( e^{i  \frac{a \hat{p}}{\hbar }} +  e^{-i  \frac{a \hat{p}}{\hbar }} \big), \nonumber 
\end{align}
where $e^{-i  \frac{a \hat{p}}{\hbar }}$ moves the particle the distance \emph{a} along the chain. The (non-normalized) eigenstates of \eqref{app:eq:H_dis} can then be picked as simultaneuous eigenstates of the discretized translation operator: 

\begin{align}
\ket{\psi_p} = \sum_{x \in \mathbb{Z}} e^{ \frac{ipax}{\hbar} } \ket{x}, \quad E^{(dis)}_p =2\beta \cos\bigg( \frac{ap}{\hbar}\bigg), \nonumber 
\end{align}
where \emph{p} is  the discretized equivalent of the momentum of the particle. In the case of a continuous space,  the free-particle Hamiltonian would be 

\begin{align}
 \hat{H}_{\text{free}} = \frac{\hat{p}^2}{2m},\label{app:eq:H_free}
\end{align}
and would have the eigenstates and eigenenergies

\begin{align}
\ket{\psi_p} =  \int^{\infty}_{-\infty} dx e^{ipx/\hbar} \ket{x}, \quad E^{(free)}_p = \frac{p^2}{2m},\nonumber 
\end{align}
where \emph{m} is the mass of the electron. The eigenstates of the two models \eqref{app:eq:H_dis} and  \eqref{app:eq:H_free} are identical up to a discretization, but the eigenenergies are not the same. To obtain a spectrum of \eqref{app:eq:H_dis} that more closely resembles that of \eqref{app:eq:H_free}, we first rewrite the energy as 

\begin{align}
E^{(\text{dis})}_p &= 2\beta \cos\bigg( \frac{a p}{\hbar}\bigg) \nonumber \\
 &=  2\beta \bigg( 1 - \frac{1}{2} \bigg( \frac{ap}{\hbar}\bigg)^2  +  O\bigg( \frac{ap}{\hbar}\bigg)^4 \bigg).  \nonumber 
\end{align}
Thus changing the Hamiltonian to 

\begin{align}
\hat{H}'_{\text{dis}} = 2\beta \mathbb{1} - \hat{H}_{dis} \nonumber 
\end{align}
would preserve the eigenstates, but yields an eigenspectrum similar of the continuum-case:

\begin{align}
E^{'(\text{dis})}_p&=\beta \bigg( \frac{ap}{\hbar}\bigg)^2 +  O\bigg( \frac{ap}{\hbar}\bigg)^4 \nonumber \\
&\approx \beta \bigg( \frac{ap}{\hbar}\bigg)^2,  \quad  \text{if $a \ll \  \frac{\hbar}{\abs{p}}$}. \nonumber 
\end{align}
That is, to increase the similarity we require that the distance between the quantum dots is much smaller than the inverse of the incoming momentum. We can further increase this similarity by setting $\beta$ to

\begin{align}
\beta \frac{a^2}{\hbar^2} = \frac{1}{2m} \quad \rightarrow \quad \beta = \frac{\hbar^2}{2ma^2}. \nonumber 
\end{align}
Thus, we can model the motion of free particles with momentum $\momentum$ using the discrete model \eqref{app:eq:H_dis} as long as \emph{a} and $\beta$ fulfill these relations.
\section{An Analysis of the Lead-Molecule Coupling}
\label{app:lead_molecule}

In this section, we will investigate the coupling between the leads and molecule. We consider the coupling term between the two subsystems to be

\begin{align}
\hat{V} &= \sum^N_{\lead = 1} \sum^M_{p=0}  \big( V_{\lead, p} \hat{c}^\dagger_{\lead,1} \hat{\tilde{a}}_{p} + V_{\lead, p}^* \hat{\tilde{a}}^\dagger_{p} \hat{c}_{\lead,1} \big), \nonumber 
\end{align}
where $\hat{\tilde{a}}_p = \hat{s}^\dagger \hat{a}_p $ and $\hat{s}$ is an operator that generate molecular orbitals with the lowest energy for the charged molecule from those of the uncharged molecule. The operator $\hat{s}$ will be useful when evaluating the coupling elements later in this section. The state of the system immediately before and after transfer is

\begin{align}
&\ket{\lead,x = 1; E^{(\NumberElectrons)}_0}   \quad &\leftarrow \quad &\text{immediately before transfer} \nonumber \\
&\ket{E^{(\NumberElectrons+1)}_0}, \ket{E^{(\NumberElectrons+1)}_1}, \hdots \quad& \leftarrow \quad &\text{immediately after transfer}, \nonumber 
\end{align}
where we have denoted the particle being at $x = 1$ along the lead $\lead$ as $\ket{\lead, x = 1}$, and the molecule being in its ground state $\ket{E^{(\NumberElectrons)}_0}$, and  -1 charged energy eigenstates as $\ket{E^{(\NumberElectrons+1)}_0}$, $\ket{E^{(\NumberElectrons+1)}_1}$ etc. In terms of Slater determinants of molecular orbitals, each labelled by an occupation number vector $\vec{k}_j \in \{0,1\}^M$, where \emph{M} is the number of molecular orbitals (or spin-orbitals) describing which orbitals are empty (0) and which are occupied (1), we can express the energy eigenstates as

\begin{align}
\ket{E^{(\NumberElectrons)}_0} &= \sum_j c^{(\NumberElectrons)}_{0,j} \ket{\vec{k}^{(\NumberElectrons)}_{j}} \nonumber \\
\ket{E^{(\NumberElectrons+1)}_g} &= \sum_j c^{(\NumberElectrons+1)}_{g,j} \ket{\vec{k}^{(\NumberElectrons+1)}_{j}}, \nonumber 
\end{align}
meaning the coupling elements take the form:

\begin{align}
\braket{E^{(\NumberElectrons+1)}_g|\hat{V}|\lead',\site = 1; E^{(\NumberElectrons)}_0} &=     \sum^N_{\lead = 1} \sum^M_{p=0}  \braket{E^{(\NumberElectrons+1)}_g|  V_{\lead, p} \hat{c}^\dagger_{\lead,1} \hat{\tilde{a}}_{p} + V_{\lead, p}^* \hat{\tilde{a}}^\dagger_{p} \hat{c}_{\lead,1} |\lead',\site = 1; E^{(\NumberElectrons)}_0  } \nonumber \\
&=  \sum_{j,j'} \big(c^{(\NumberElectrons+1)}_{g,j}\big)^*c^{(\NumberElectrons)}_{0,j'}   \sum^N_{\lead = 1} \sum^M_{p=0} V_{\lead, p}^* \delta_{\lead,\lead'}  \braket{\vec{k}^{(\NumberElectrons+1)}_{j}|  \hat{\tilde{a}}^\dagger_{p} | \vec{k}^{(\NumberElectrons)}_{j'}  } \label{app:eq:V_bracket} \\
&=  \sum_{j,j'} \big(c^{(\NumberElectrons+1)}_{g,j}\big)^*c^{(\NumberElectrons)}_{0,j'}   \sum^N_{\lead = 1} \sum^M_{p=0}  V_{\lead, p}^*  \delta_{\lead,\lead'} \delta_{\norm{\vec{k}^{(\NumberElectrons +1)}_j-\vec{k}^{(\NumberElectrons )}_{j'}},1} \delta_{\vec{p},\vec{k}^{(\NumberElectrons +1)}_j-\vec{k}^{(\NumberElectrons )}_{j'}} \nonumber \\
&= \sum_{\norm{\vec{k}^{(\NumberElectrons +1)}_j-\vec{k}^{(\NumberElectrons )}_{j'}} = 1}   \big(c^{(\NumberElectrons+1)}_{g,j}\big)^*c^{(\NumberElectrons)}_{0,j'}   V_{\lead', \vec{k}^{(\NumberElectrons +1)}_j-\vec{k}^{(\NumberElectrons )}_{j'}}^*, \nonumber 
\end{align}
For example, if the Jordan-Wigner mapping is used to map occupation-numbers to computational basis states, and if  $\ket{\vec{k}^{(3)}_{j}} = \ket{1110} $ and  $\ket{\vec{k}^{(2)}_{j'}} = \ket{0011}$, then $\norm{\vec{k}^{(3)}_j-\vec{k}^{(2)}_{j'}} = 3$, and the term vanishes due to mismatch of the occupation number vectors. In fact only terms that satisfy $\norm{\vec{k}^{(\NumberElectrons +1)}_j-\vec{k}^{(\NumberElectrons )}_{j'}} = 1$ can be non-zero. We have used a slight abuse of notation: In the main text, we denoted the coupling elements as $V_{\lead,p}$, where $p \in \mathbb{Z}$, but here we use $V_{\lead', \vec{k}_j-\vec{k}_{j'} }$? The reason is that we for brevity usually number the elements of the set $\{\vec{k}^{(\NumberElectrons +1)}_j-\vec{k}^{(\NumberElectrons)}_{j'} :\norm{\vec{k}^{(\NumberElectrons +1)}_j-\vec{k}^{(\NumberElectrons)}_{j'}} = 1\} $ to allow for a more compact/convenient notation. Note, if we did not include the operator $\hat{s}$ in $\hat{\tilde{a}}_p = \hat{s}^\dagger \hat{a}_p $, then the overlap in Eq. \eqref{app:eq:V_bracket} would not be as simple as

\begin{align}
\braket{\vec{k}^{(\NumberElectrons+1)}_{j}|  \hat{a}^\dagger_{p} | \vec{k}^{(\NumberElectrons)}_{j'}  } \neq\delta_{\norm{\vec{k}^{(\NumberElectrons +1)}_j-\vec{k}^{(\NumberElectrons )}_{j'}},1} \delta_{\vec{p},\vec{k}^{(\NumberElectrons +1)}_j-\vec{k}^{(\NumberElectrons )}_{j'}}. \nonumber 
\end{align}
For example, consider a system of four molecular orbitals $\ket{\chi_0,\chi_1,\chi_2,\chi_3}$ and two electrons. Let $\chi_p^{(\NumberElectrons+1)}$ ($\chi_p^{(\NumberElectrons)}$) be the \emph{p}'th molecular orbital for the charged (neutral) molecule. Without the operator $\hat{s}$, the following overlap evaluates to 

\begin{align}
 \braket{\vec{k}^{(\NumberElectrons+1)}_{j}|\hat{a}^\dagger_{1}|\vec{k}^{(\NumberElectrons)}_{j'} }&=\braket{1110|\hat{a}^\dagger_{1}|0011} \nonumber \\
 &=\braket{\chi_0^{(3)}\chi_2^{(3)}|\chi_2^{(2)}\chi_3^{(2)}} \nonumber \\
 &\neq 0 \nonumber 
\end{align}
since in general $\braket{\chi_p^{(3)}|\chi_q^{(2)}} \neq \delta_{pq}$.  Including the operator $\hat{s}$, we obtain 

\begin{align}
\braket{1110|\hat{a}^\dagger_{1}\hat{s} |0011} &=\braket{\chi_0^{(3)}\chi_2^{(3)}| \hat{s} |\chi_2^{(2)}\chi_3^{(2)}} \nonumber \\
   &=\braket{\chi_0^{(3)}\chi_2^{(3)}|\chi_2^{(3)}\chi_3^{(3)}} \nonumber \\
 & = 0 \nonumber 
\end{align}
since $\braket{\chi_p^{(3)}|\chi_q^{(3)}} = \delta_{pq}$. That is, the operator $\hat{s}$ moves the electrons and shuffles the molecular orbitals $\{\chi_p^{(2)}\}$ to generate the molecular orbitals $\{\chi_p^{(3)}\}$. 

To simplify the coupling, we assume that we can separate the leads and molecular orbitals indices as

\begin{align}
V_{\lead,p} = V_{\lead} V_{p},\label{app:eq_V_int_1}
\end{align}
Furthermore, we assume that $V_{p}$ equals the absolute square of the molecular orbital which takes part in the transfer process, integrated over a region close to the leads:

\begin{align}
V_{p} =   \int_{\mathcal{V}} d\vec{r} \hspace{0.1cm} \abs{\chi_{p}(\vec{r})}^2, \label{app:eq_V_int}
\end{align}
where $\chi_{p}(\vec{r})$ is the \emph{p}'th molecular orbital of the uncharged molecule. It seems reasonable that if the molecular orbital density is very small near the lead, then consequently the coupling strength through this particular orbital should be small as well, hence the expression in Eq. \eqref{app:eq_V_int}.

\section{The One-Particle Scattering Matrix}
\label{app:One_Particle_Scattering_Matrix}
As mentioned in the main text, we look for a solution of the form 

\begin{align}
\hat{H} \ket{\text{sc}_{\lead}(\momentum)} & = \big(\hat{H}_{\text{leads}} + \hat{H}_{\text{int}} + \hat{V} \big)\ket{\text{sc}_{\lead}(\momentum)} \nonumber \\
&= E \ket{\text{sc}_{\lead}(\momentum)}, \label{app:eq:eign}
\end{align}
where $\ket{\text{sc}_{\lead}(\momentum)}$ is the one-particle scattering ansatz \eqref{eq:one_particle_sc}, and the Hamiltonian terms are given in \eqref{eq:dis_hamil}-\eqref{eq:V_hamil}. The scattering ansatz is an eigenstate of $\hat{H}$ for some appropriate amplitudes $w_{\lead,g}$ and $S_{\lead, \lead'}$. To find these amplitudes, we first evaluate each term in \eqref{app:eq:eign}:

\begin{align}
\hat{H}_{\text{leads}}\ket{\text{sc}_{\lead}(\momentum)} &= \bigg(2\beta-2\beta \cos\bigg( \frac{\momentum a}{\hbar} \bigg)\bigg) \sum^N_{\lead' = 1} \sum^{\infty}_{x = 1} \big(e^{-i \frac{\momentum a x }{\hbar}} \delta_{\lead,\lead'} + S_{\lead,\lead'} e^{i \frac{\momentum a x }{\hbar}} \big)\ket{\lead', x; E^{(\NumberElectrons)}_0} \nonumber \\
&+\beta \sum^N_{\lead ' = 1}  \big(\delta_{\lead,\lead'} + S_{\lead, \lead'}\big)\ket{\lead',1;E^{(\NumberElectrons)}_0}  \nonumber \\[0.3cm]
\hat{H}_{\text{int}}\ket{\text{sc}_{\lead}(\momentum)} &= E^{(\NumberElectrons)}_0 \sum^N_{\lead' = 1} \sum^{\infty}_{x = 1} \big(e^{-i \frac{\momentum a x }{\hbar}} \delta_{\lead,\lead'} + S_{\lead,\lead'} e^{i \frac{\momentum a x }{\hbar}} \big)\ket{\lead', x; E^{(\NumberElectrons)}_0} + \sum^G_{g = 0} E_g^{(\NumberElectrons+1)} w_{\lead,g} \ket{E^{(\NumberElectrons+1)}_g}. \nonumber 
\end{align}
Using the results of section \ref{app:kinetics}: 

\begin{align}
\hat{H}_{\text{leads}}\ket{\text{sc}_{\lead}(\momentum)} &\approx \frac{p^2}{2m} \sum^N_{\lead' = 1} \sum^{\infty}_{x = 1} \big(e^{-i \frac{\momentum a x }{\hbar}} \delta_{\lead,\lead'} + S_{\lead,\lead'} e^{i \frac{\momentum a x }{\hbar}} \big)\ket{\lead', x; E^{(\NumberElectrons)}_0} \nonumber \\ 
&+ \frac{\hbar^2}{2ma^2}\sum^N_{\lead ' = 1}  \big(\delta_{\lead',\lead} + S_{\lead, \lead'}\big)\ket{\lead',1;E^{(\NumberElectrons)}_0}.  \nonumber
\end{align}
In order to evaluate the coupling term, we assume the electrons are tightly bound to the  uncharged molecule. Specifically, we assume that the electrons initially bound to the molecule do not exit the molecule to travel along the leads --- an assumption which corresponds to omitting contributions of the form $\hat{\tilde{a}}_p \ket{\lead, x;E^{(\NumberElectrons)}_0}$. We find

\begin{align}
\hat{V} \ket{\text{sc}_{\lead}(\momentum)} 
&= \sum^N_{\lead' = 1} \sum^M_{p = 0}  \sum^G_{g = 0}  V_{\lead',p}  w_{\lead,g} \hat{c}^\dagger_{\lead',1} \hat{\tilde{a}}_p \ket{E^{(\NumberElectrons+1)}_g}\nonumber \\
&+\sum^N_{\lead',\lead'' = 1}\sum^\infty_{x= 1}  \sum^M_{p=0}  V^*_{\lead'', p}  \big( e^{-i \frac{p a x}{\hbar}}  \delta_{\lead, \lead'} +e^{i \frac{p a x}{\hbar}}  S_{\lead, \lead'}  \big)  \hat{\tilde{a}}^\dagger_p  \hat{c}_{\lead'',1} \ket{\lead', x; E^{(\NumberElectrons)}_0} \nonumber \\
&= \sum^N_{\lead' = 1} \sum^M_{p = 0}  \sum^G_{g=0}  V_{\lead',p}  w_{\lead,g} \hat{c}^\dagger_{\lead',1} \hat{\tilde{a}}_p \ket{E^{(\NumberElectrons+1)}_g}\nonumber \\
&+\sum^N_{\lead' = 1} \sum^M_{p=0}  V^*_{\lead', p}  \big( e^{-i \frac{p a }{\hbar}}  \delta_{\lead, \lead'} +e^{i \frac{p a }{\hbar}}  S_{\lead, \lead'} \big)  \hat{\tilde{a}}_p^\dagger \ket{E^{(\NumberElectrons)}_0}.  \nonumber 
\end{align}
We can expand the states in the occupation number basis and insert $\mathbb{1}=\sum_E \ket{E}\bra{E}$:

\begin{align}
\hat{\tilde{a}}^\dagger_p \ket{E^{(\NumberElectrons)}_0} &= \sum^G_{g=0}     \braket{E^{(\NumberElectrons+1)}_g|\hat{\tilde{a}}^\dagger_p |E^{(\NumberElectrons)}_0 } \ket{E^{(\NumberElectrons+1)}_g} \nonumber \\
 \hat{\tilde{a}}_p\ket{E^{(\NumberElectrons+1)}_g} &= \sum^L_{l=0}     \braket{E^{(\NumberElectrons)}_l|\hat{\tilde{a}}_p |E^{(\NumberElectrons+1)}_g } \ket{E^{(\NumberElectrons)}_l} \label{app:eq:ENplus1}
\end{align}
Putting it all together, we obtain 

\begin{align}
\hat{H}\ket{\text{sc}_{\lead}(\momentum)}
&= \bigg( \frac{p^2}{2m} +E^{(\NumberElectrons)}_0 \bigg) 
\sum^N_{\lead' = 1} \sum^{\infty}_{x = 1}  \big(e^{-i \frac{\momentum a x }{\hbar}} \delta_{\lead,\lead'} + S_{\lead,\lead'} e^{i \frac{\momentum a x }{\hbar}} \big)\ket{\lead', x; E^{(\NumberElectrons)}_0} \label{app:eq:sca1} \\
&+ \frac{\hbar^2}{2ma^2}\sum^N_{\lead ' = 1}\big(\delta_{\lead,\lead'} + S_{\lead, \lead'}\big)\ket{\lead',1;E^{(\NumberElectrons)}_0} \\
&+ \sum^G_{g = 0} E_g^{(\NumberElectrons+1)} w_{\lead,g} \ket{E^{(\NumberElectrons+1)}_g} \label{app:eq:sca2} \\
&+ \sum^N_{\lead' = 1} \sum^M_{p = 0}  \sum^G_{g = 0} \sum^L_{l = 0} V_{\lead',p}  w_{\lead,g} \braket{E^{(\NumberElectrons)}_l|\hat{\tilde{a}}_p |E^{(\NumberElectrons+1)}_g } \ket{\lead',1;E^{(\NumberElectrons)}_l} \label{app:eq:sca3} \\
&+\sum^N_{\lead' = 1} \sum^M_{p=0} \sum^G_{g =0}  V^*_{\lead', p} \braket{E^{(\NumberElectrons+1)}_g|\hat{\tilde{a}}^\dagger_p |E^{(\NumberElectrons)}_0 }  \big( e^{-i \frac{ p a }{\hbar}}  \delta_{\lead, \lead'} +e^{i \frac{ p a }{\hbar}}  S_{\lead, \lead'}  \big) \ket{E^{(\NumberElectrons + 1)}_g}. \label{app:eq:sca4}
\end{align}
The right-hand side of the Schr\"odinger equation \eqref{app:eq:eign} simply becomes

\begin{align}
E \ket{\text{sc}_{\lead}(\momentum)} =E \sum^N_{\lead' = 1} \sum^{\infty}_{x = 1} \big(e^{-i \frac{\momentum a x }{\hbar}} \delta_{\lead,\lead'} + S_{\lead,\lead'} e^{i \frac{\momentum a x }{\hbar}} \big) \ket{\lead', x; E^{(\NumberElectrons)}_0} +E \sum^G_{g=0}  w_{\lead,g} \ket{E^{(\NumberElectrons+1)}_g}. \label{app:eq:sca5}
\end{align}
Considering the terms in Eqs. \eqref{app:eq:sca1} and  \eqref{app:eq:sca5}, we let $E = \momentum^2/2m +E^{(\NumberElectrons)}_0$. We have then solved the Schr\"odinger equation \eqref{app:eq:eign} with our particular form of state (and energy) if and only if we can make the amplitudes in \eqref{app:eq:sca5} match those in \eqref{app:eq:sca1}-\eqref{app:eq:sca4}. The number of unknowns are $N + (G+1)$, and the number of equations are $N\times(L+1) + (G+1)$. To overcome the problem of an overdetermined system, we will only include the ground state in \eqref{app:eq:ENplus1}. This corresponds to assuming that the molecular state is not changed by the scattering. Note that this means we restrict our description to elastic scattering events because the molecular energy is conserved, hence so is the electron energy, and thus only the direction of travel for the electron is changed when it scattered through the molecule. To solve the equations, we must pick our parameters so that

\begin{align}
&\frac{\hbar^2}{2ma^2} \big(\delta_{\lead,1} + S_{\lead, 1}\big) +\sum^M_{p=0} \sum^G_{g=0}  V_{1,p}w_{\lead, g} \braket{E^{(\NumberElectrons)}_0|\hat{\tilde{a}}_p |E^{(\NumberElectrons+1)}_g } = 0 \label{app:eq:S_eq_1} \\
\vdots \nonumber \\
&\frac{\hbar^2}{2ma^2} \big(\delta_{\lead,N} + S_{\lead, N}\big) +\sum^M_{p=0} \sum^G_{g=0}  V_{N,p}w_{\lead, g}\braket{E^{(\NumberElectrons)}_0|\hat{\tilde{a}}_p |E^{(\NumberElectrons+1)}_g } = 0   \label{app:eq:S_eq_2} \\[0.3cm]   
& E_0^{(\NumberElectrons+1)} w_{\lead,0} +\sum^N_{\lead' = 1} \sum^M_{p=0}  V^*_{\lead', p} \braket{E^{(\NumberElectrons+1)}_g|\hat{\tilde{a}}^\dagger_p |E^{(\NumberElectrons)}_0 }  \big( e^{-i \frac{ p a }{\hbar}}  \delta_{\lead, \lead'} +e^{i \frac{ p a }{\hbar}}  S_{\lead, \lead'}  \big)  = \big( \frac{\momentum^2}{2m} +E^{(\NumberElectrons)}_0 \big) w_{\lead,0} \nonumber \\
\vdots \nonumber \\
& E_G^{(\NumberElectrons+1)} w_{\lead,G} +\sum^N_{\lead' = 1} \sum^M_{p=0}  V^*_{\lead', p} \braket{E^{(\NumberElectrons+1)}_g|\hat{\tilde{a}}^\dagger_p |E^{(\NumberElectrons)}_0 }  \big( e^{-i \frac{ p a }{\hbar}}  \delta_{\lead, \lead'} +e^{i \frac{ p a }{\hbar}}  S_{\lead, \lead'}  \big)  = \big( \frac{\momentum^2}{2m} +E^{(\NumberElectrons)}_0 \big) w_{\lead,G}.\label{app:eq:S_eq_N} 
\end{align}
Notice

\begin{align}
\braket{E^{(\NumberElectrons+1)}_g|\hat{\tilde{a}}^\dagger_p |E^{(\NumberElectrons)}_0 }^* = \braket{E^{(\NumberElectrons)}_0|\hat{\tilde{a}}_p |E^{(\NumberElectrons+1)}_g }. \nonumber 
\end{align}
In matrix representation, we obtain 

\begin{align}
\begin{pmatrix}
       0 &B^\dagger\\[0.3em]
      B& D   
     \end{pmatrix} \begin{pmatrix}
        e^{-i \frac{ p a }{\hbar}}  +e^{i \frac{ p a }{\hbar}}  S \\[0.3em]
      \Psi  
     \end{pmatrix}  +\frac{\hbar^2}{2ma^2} \begin{pmatrix}
        1 + S\\[0.3em]
      0   
     \end{pmatrix} = \bigg( \frac{\momentum^2}{2m} +E^{(\NumberElectrons)}_0\bigg)\begin{pmatrix}
        0 \\[0.3em]
       \Psi     
     \end{pmatrix}, \label{eq:app:scattering_eq}
\end{align}
where

\begin{gather}
S = \begin{pmatrix}
  S_{1,1} & S_{2,1} & \cdots & S_{N,1} \\
  S_{1,2} & S_{2,2} & \cdots & S_{N,2} \\
  \vdots  & \vdots  & \ddots & \vdots  \\
  S_{1,N} & S_{2,N} & \cdots & S_{N,N} 
 \end{pmatrix}  \quad \Psi = \begin{pmatrix}
  w_{1,0} & w_{2,0} & \cdots & w_{N,0} \\
  w_{1,1} & w_{2,1} & \cdots & w_{N,1} \\
  \vdots  & \vdots  & \ddots & \vdots  \\
  w_{1,G} & w_{2,G} & \cdots & w_{N,G} 
 \end{pmatrix} \nonumber \\[0.3cm]  D =   \begin{pmatrix}
E_0^{(\NumberElectrons +1)} & \cdots &  0 \\
  \vdots   & \ddots & \vdots  \\
  0& \cdots &  E_G^{(\NumberElectrons +1)}
 \end{pmatrix}    \nonumber \\[0.3cm] 
B =   \begin{pmatrix}
 \sum^M_{p=0}  V^*_{1, p}\braket{E^{(\NumberElectrons + 1)}_0|\hat{\tilde{a}}^\dagger_p|E^{(\NumberElectrons)}_{0}} & \cdots &  \sum^M_{p=0}   V^*_{N, p} \braket{E^{(\NumberElectrons + 1)}_0|\hat{\tilde{a}}^\dagger_p|E^{(\NumberElectrons)}_{0}}  \\
  \vdots   & \ddots & \vdots  \\
  \sum^M_{p=0} V^*_{1, p} \braket{E^{(\NumberElectrons + 1)}_G|\hat{\tilde{a}}^\dagger_p|E^{(\NumberElectrons)}_{0}}  & \cdots & \sum^M_{p=0}   V^*_{N, p} \braket{E^{(\NumberElectrons + 1)}_G|\hat{\tilde{a}}^\dagger_p|E^{(\NumberElectrons)}_{0}} 
 \end{pmatrix}. \nonumber 
\end{gather}
Equation \eqref{eq:app:scattering_eq} is the same result as in Refs. \cite{childs_levinsons_2011, childs_levinsons_2012} for scattering through quantum graphs. The matrix elements in the coupling matrix \emph{B} can be evaluated using the approximation described in section \ref{app:lead_molecule}. The lower part of \eqref{eq:app:scattering_eq} gives: 

\begin{align}
\Psi = \frac{1}{\momentum^2/2m +E^\NumberElectrons_0 - D }  \big( e^{-i \frac{ p a }{\hbar}} B +e^{i \frac{ p a }{\hbar}} BS \big), \nonumber 
\end{align}
and the upper part gives the scattering matrix: 

\begin{align}
S(e^{i \frac{ p a }{\hbar}}) = - Q(e^{i \frac{ p a }{\hbar}})^{-1} Q(e^{-i \frac{ p a }{\hbar}}), \label{eq:app:S_matrix}
\end{align}
where

\begin{align}
Q(e^{i \frac{ p a }{\hbar}}) =  \frac{\hbar^2}{2ma^2} + e^{i \frac{ p a }{\hbar}}  B^\dagger \frac{1}{\momentum^2/2m +E^{(\NumberElectrons)}_0 - D } B. \nonumber 
\end{align}
The scattering matrix is unitary:

\begin{align}
S(e^{i \frac{ p a }{\hbar}})^\dagger &=   -Q(e^{-i \frac{ p a }{\hbar}})^\dagger \big( Q(e^{i \frac{ p a }{\hbar}})^{-1} \big)^{\dagger} \nonumber \\
&= - Q(e^{i \frac{ p a }{\hbar}})  Q(e^{-i \frac{ p a }{\hbar}})^{-1} \nonumber \\
&= -   Q(e^{-i \frac{ p a }{\hbar}})^{-1}  Q(e^{i \frac{ p a }{\hbar}})  \nonumber \\
&=S(e^{i \frac{ p a }{\hbar}})^{-1},  \nonumber
\end{align}
since $Q(e^{-i \frac{ p a }{\hbar}}) = Q(e^{i \frac{ p a }{\hbar}})^\dagger$ and $[Q(e^{i \frac{ p a }{\hbar}}), Q(e^{-i \frac{ p a }{\hbar}})] = 0$ meaning $[Q(e^{i \frac{ p a }{\hbar}}), Q(e^{-i \frac{ p a }{\hbar}})^{-1}] = 0$. The unitarity of the \emph{S}-matrix was also pointed out in Ref.~\cite{childs_levinsons_2012}. Having established the \emph{S}-matrix, we will now look at the dynamics, specifically the total distance traveled in some amount of time \emph{t}. This is important in order to control the timing of how the electrons propagate. For instance, if the scattering process significantly slows the electron down compared to an electron propagating on a lead with no molecule,  then the time differences may affect the computation. Ideally, the effective length, and hence the traversal time, should be the same with or without a molecule attached between the leads. To analyze this, we use the same result as in Ref. \cite{childs_universal_2009}, which computes the effective length for quantum graphs, a length that in our case corresponds to the effective distance traveled within the molecule. Assume we want to travel from a point \emph{x} on one chain $\lead$, through the molecule and into a point \emph{y} on a different chain $\lead'$. Since $\braket{\text{sc}_{\lead'}(\momentum')|\text{sc}_{\lead}(\momentum)} = 2\pi \delta_{\lead',\lead} \delta(\momentum'-\momentum)$ [proof for normalization can be found in Ref. \cite{childs_universal_2013}, section S1],  the dynamics of the system can be written as 

\begin{align}
\braket{\lead',y|e^{-i\hat{H}t/\hbar}|\lead,x} &\approx\sum^N_{\lead'' = 1} \int^{\infty}_{0} \frac{d\momentum}{2\pi} e^{-i t \big(\frac{\momentum^2}{2m} + E^{(\NumberElectrons)}_0 \big)/\hbar } \braket{\lead',y | \text{sc}_{\lead''}(\momentum)} \braket{\text{sc}_{\lead''}(\momentum)|\lead,x} \nonumber \\
&= \int^{\infty}_{0} \frac{d\momentum}{2\pi} e^{-i t \big(\frac{\momentum^2}{2m} + E^{(\NumberElectrons)}_0 \big)/\hbar } \bigg( S_{\lead,\lead'}(\momentum) e^{i \frac{ p a }{\hbar} (x + y)} + S^*_{\lead',\lead}(\momentum) e^{-i \frac{ p a }{\hbar} (x + y)} \bigg), \label{eq:app:one_par_dyn}
\end{align}
where the approximation sign comes from the results in section \ref{app:kinetics}, i.e., $ \beta(2 - 2 \cos(a\momentum/\hbar)) \approx \momentum^2/2m$. We neglected the bound states (i.e. the normalized states) in Eq. \eqref{eq:app:one_par_dyn}, since these go to zero for large values of \emph{x} and \emph{y}~\cite{childs_universal_2009}. The integral \eqref{eq:app:one_par_dyn} is dominated by those values where the momentum-derivative of the phase vanishes~\cite{wong_asymptotic_2001, childs_universal_2009}, and it can be shown that the second term has no stationary points for $t >0$. Thus, only stationary point is the one where $ -\frac{d}{d\momentum} \frac{t}{\hbar}\frac{\momentum^2}{2m} + \frac{a}{\hbar}(x+y) + \frac{d}{d\momentum} \text{Arg}( S_{\lead,\lead'})) = 0 $, and we find 

\begin{align}
a(x + y) +  \ell_{\lead,\lead'} = v t, \label{eq:app:effec_length} 
\end{align}
where

\begin{align}
\ell_{\lead,\lead'} &= \hbar \frac{d}{d\momentum} \text{Arg}( S_{\lead,\lead'})    \label{eq:app:effec_length_1}
\end{align}
is the effective length (here we follow the same notation as in Ref. \cite{childs_universal_2009}) and $v = \momentum / m$ is the speed of the incoming electron with kinetic energy $\momentum^2/2m$. Equation \eqref{eq:app:effec_length} came from evaluating the dynamics in Eq. \eqref{eq:app:one_par_dyn}, and it tells us that the total distance traveled in some amount of time \emph{t} to go from point \emph{x} on one chain to point \emph{y} on another is $a(x+y)$ plus an additional non-trivial term $\ell_{\lead,\lead'} $ which is the effective distance traveled within the molecule.

\subsection{Transmission Phase Through a Linear Chain of Quantum Dots}
\label{app:example_lead_effective_length}
In this example, we will find an expression for the transmission amplitude and phase through a straight lead. This result was already pointed out as a footnote in Ref. \cite{childs_universal_2009}, and here we give a derivation of their result. Assume that we have split the lead into 3 sub-leads, as depicted in Fig. \ref{fig:linear_chain}, and that we label the states of each sub-lead as $\ket{\lead = 1, \site}$, $\ket{g}$ and $\ket{\lead =2, \site}$, respectively. The Hamiltonian can then be divided as follows:

\begin{figure}[H] 
\centering  
\includegraphics[width=0.6\textwidth]{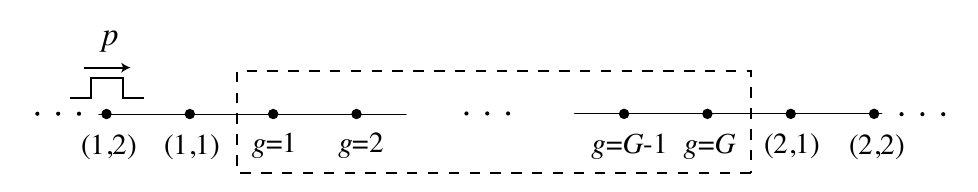}
\caption{ Schematic depiction of a linear chain of dots each representing a quantum state coupled to its two nearest neighbours. The quantum dots along the leads are labeled as $(\lead,\site)$, where $\lead$ and $\site$ indicate the lead and site, respectively.}
\label{fig:linear_chain}
\end{figure}

\begin{align}
    \hat{H} = \hat{H}_{\text{leads}} + \hat{H}_G + \hat{V}, \label{eq:app:hamil_straight_line}
\end{align}
where 

\begin{align}
 \hat{H}_{\text{leads}} &= \sum^2_{\lead =1} \sum^\infty_{\site = 1} \bigg(  2 \beta \hat{c}^\dagger_{\lead,\site} \hat{c}_{\lead,\site}  -\beta \big(\hat{c}^\dagger_{\lead,\site} \hat{c}_{\lead,\site + 1} + \hat{c}^\dagger_{\lead,\site + 1}\hat{c}_{\lead,\site}   \big) \bigg), \nonumber \\
 \hat{H}_G & =  \sum^{G-1}_{g = 1} \bigg(  2 \beta \hat{c}^\dagger_{g} \hat{c}_{g}  -\beta \big(\hat{c}^\dagger_{g} \hat{c}_{g + 1} + \hat{c}^\dagger_{g + 1} \hat{c}_{g}\big) \bigg) +  2 \beta \hat{c}^\dagger_{G} \hat{c}_{G}   \nonumber \\
 \hat{V} &= -\beta \big(\hat{c}^\dagger_{\lead = 1,\site = 1} \hat{c}_{g = 1} + \hat{c}^\dagger_{g = 1}  \hat{c}_{\lead = 1,\site = 1}\big)  -\beta \big(\hat{c}^\dagger_{\lead = 2,\site = 1} \hat{c}_{g = G} + \hat{c}^\dagger_{g = G} \hat{c}_{\lead = 2,\site = 1}\big). \nonumber 
\end{align}
As seen above, the approach outlined by Childs \emph{et al.}~\cite{childs_levinsons_2011,childs_levinsons_2012, childs_universal_2009, childs_universal_2013} for finding eigenstates to describe the scattering of particles on the central \emph{G}-vertex graph is to look for solutions of the following form: 

\begin{align}
\ket{\text{sc}_{\lead}(\momentum)} = \sum^2_{\lead' = 1} \sum^\infty_{x = 1} \big( e^{-i \frac{\momentum a x}{\hbar}} \delta_{\lead, \lead'} + e^{i \frac{\momentum a x}{\hbar}} S_{\lead, \lead'}(\momentum) \big) \ket{\lead',x} + \sum^G_{g=1} w_{\lead,g}(\momentum) \ket{g}. \label{eq:app:childs_scattering}
\end{align}
The scattering state \eqref{eq:app:childs_scattering} differs from \eqref{eq:one_particle_sc} by expressing the scattering region in the site basis and not in the eigenbasis, however, one can easily go from one basis to another:

\begin{align}
\sum^G_{g=1} w_{\lead,g} \ket{g} &=     \sum^G_{g=1} w_{\lead,g} \sum_{\alpha} \braket{E_\alpha|g} \ket{E_\alpha} = \sum_\alpha \tilde{w}_{\lead,\alpha} \ket{E_\alpha}. \nonumber
\end{align}
If we let the Hamiltonian \eqref{eq:app:hamil_straight_line} act on the scattering state, we obtain:

\begin{align}
\hat{H} \ket{\text{sc}_{\lead}(\momentum)} &= \bigg(2\beta  -  2\beta\cos\bigg(\frac{\momentum a }{\hbar} \bigg)\bigg) \sum^2_{\lead' = 1} \sum^\infty_{x = 1} \big( e^{-i \frac{\momentum a x}{\hbar}} \delta_{\lead, \lead'} + e^{i \frac{\momentum a x}{\hbar}} S_{\lead, \lead'}(\momentum) \big) \ket{\lead',x} \nonumber \\
&+ \beta \sum^2_{\lead' =1} \bigg(\delta_{\lead,\lead'} + S_{\lead,\lead'}\bigg)\ket{\lead',1} \label{eq:app:linear_chain_1}  \\
&+2\beta \sum^{G}_{g=1} w_{\lead,g} \ket{g} -\beta \sum^G_{g = 2} w_{\lead,g} \ket{g-1} - \beta \sum^{G-1}_{g = 1} w_{\lead,g} \ket{g+1} \nonumber \\
&-\beta  w_{\lead,1} \ket{1, 1} - \beta \bigg( e^{-i \frac{\momentum a}{\hbar}} \delta_{\lead, 1} + e^{i \frac{\momentum a }{\hbar}} S_{\lead, 1}\bigg) \ket{g = 1} \nonumber \\
&- \beta  w_{\lead,G} \ket{2,1} - \beta \bigg( e^{-i \frac{\momentum a }{\hbar}} \delta_{\lead, 2} + e^{i \frac{\momentum a  }{\hbar}} S_{\lead, 2}\bigg) \ket{g = G}. \nonumber
\end{align}
Writing up the analogues of Eqs.  \eqref{app:eq:S_eq_1}-\eqref{app:eq:S_eq_N}  for this system, we can solve for the amplitudes $\{  S_{\lead, \lead'}(\momentum), w_{\lead,g}(\momentum) \}$ such that the scattering state \eqref{eq:app:childs_scattering} is an eigenstate to the Hamiltonian \eqref{eq:app:hamil_straight_line}. Considering the terms in Eq. \eqref{eq:app:linear_chain_1}, we let $E = 2\beta  -  2\beta\cos(\frac{\momentum a }{\hbar})$. To solve the equations, we must pick our parameters so that

\begin{align}
&\delta_{\lead,1} + S_{\lead,1} - w_{\lead,1} = 0  \nonumber \\
&\delta_{\lead,2} + S_{\lead,2} -  w_{\lead,G}  = 0    \nonumber  \\
&w_{\lead,2} + e^{-i \frac{\momentum a}{\hbar}} \delta_{\lead, 1} + e^{i \frac{\momentum a }{\hbar}} S_{\lead, 1}=  \bigg( e^{\frac{\momentum a }{\hbar}} +  e^{-\frac{\momentum a }{\hbar}} \bigg) w_{\lead,1}\label{app:eq:linear_chain_eqs_3} \\
&w_{\lead,G-1} + e^{-i \frac{\momentum a }{\hbar}} \delta_{\lead, 2} + e^{i \frac{\momentum a  }{\hbar}} S_{\lead, 2} = \bigg( e^{\frac{\momentum a }{\hbar}} +  e^{-\frac{\momentum a }{\hbar}} \bigg) w_{\lead,G}  \label{app:eq:linear_chain_eqs_4} \\
&w_{\lead,g+1} + w_{\lead,g-1} = \bigg( e^{\frac{\momentum a }{\hbar}} +  e^{-\frac{\momentum a }{\hbar}} \bigg) w_{\lead,g} \quad   \forall g \in \{2,3,\hdots, G-1\}. \label{app:eq:linear_chain_eqs_5}
\end{align}
Firstly, we can solve for the internal amplitudes as a function of the scattering coefficients. A reasonable guess of a solution is that the plane-wave structure of the wave function does not change when you move across the internal graph:

\begin{align}
w_{n,g}(\momentum) &= e^{i \frac{\momentum a (g-1) }{\hbar}} \delta_{\lead, 1} +  e^{-i \frac{\momentum a (g-1) }{\hbar}} S_{\lead, 1} & \forall g \in \{1,2,\hdots, G\}. \label{app:eq:linear_chain:w} 
\end{align}
Equation \eqref{app:eq:linear_chain:w} fulfills Eqs. \eqref{app:eq:linear_chain_eqs_3} and \eqref{app:eq:linear_chain_eqs_5}, and Eq. \eqref{app:eq:linear_chain_eqs_4} if $S_{11} = 0$ and $S_{2,1} = e^{i \frac{\momentum a (G-1) }{\hbar}} $. The remaining scattering coefficients can then be solved for yielding:

\begin{align}
S_{1,1} &= 0  \nonumber \\
S_{1,2}(\momentum) &= e^{i \frac{\momentum a (G-1) }{\hbar}} \nonumber \\
S_{2,1}(\momentum) &= e^{i \frac{\momentum a (G-1) }{\hbar}} \nonumber \\
S_{2,2}&=0. \nonumber 
\end{align}
Thus we have perfect transmission and a phase of $e^{i \frac{\momentum a (G-1) }{\hbar}}$ applied in the process. The effective length Eq. \eqref{eq:app:effec_length_1} turns out to be $a(G-1)$.

\subsection{Separating of Lead's and Molecular Orbital's indices for the Coupling Elements}
\label{app:example_independent_from_leads}
As discussed in section \ref{app:lead_molecule}, we assume that we can separate the lead's and molecular orbital's indices as

\begin{align}
V_{\lead,p} = V_{\lead} V_{p}. \label{eq:app:V:constraint}
\end{align}
That is to say that the rows in the coupling matrix \emph{B} (just below Eq. \eqref{eq:app:scattering_eq}) are identical for each lead up to a constant, and that each matrix element can  be written as 

\begin{align}
\sum^M_{p = 0}  V_{\lead,p}^* \braket{E^{(\NumberElectrons +1)}_g | \hat{\tilde{a}}^\dagger_p | E^{(\NumberElectrons)}_0} = V^*_{\lead} \sum^M_{p = 0}  V_{p}^* \braket{E^{(\NumberElectrons +1)}_g | \hat{\tilde{a}}^\dagger_p | E^{(\NumberElectrons)}_0} \equiv V_{\lead}^* c_{g}, \nonumber
\end{align}
where $c_g \equiv  \sum^M_{p = 0}  V_{p}^* \braket{E^{(\NumberElectrons +1)}_g | \hat{\tilde{a}}^\dagger_p | E^{(\NumberElectrons)}_0} $. Given the assumption in \eqref{eq:app:V:constraint}, we want to find a general expression for the scattering coefficients. Equations \eqref{app:eq:S_eq_1} and \eqref{app:eq:S_eq_2} reads

\begin{align}
&\frac{\hbar^2}{2ma^2} \big(\delta_{\lead,\lead'} + S_{\lead, \lead'}(\momentum)\big) + \frac{V_{\lead'}}{K_{\lead}(\momentum)}   = 0    \nonumber \\
\Rightarrow \quad &\abs{S_{\lead,\lead'}(\momentum)}^2 =  \bigg | \frac{2ma^2}{\hbar^2}  \frac{V_{\lead'}}{K_{\lead}(\momentum)} + \delta_{\lead,\lead'} \bigg |^2, \nonumber
\end{align}
where we have defined $1/K_\lead(\momentum) \equiv \sum^G_{g=0} w_{\lead,g}(\momentum)c^*_{g}$. From the unitarity of the \emph{S}-matrix, we can write 

\begin{align}
1 &= \sum^N_{\lead' = 1} \abs{S_{\lead,\lead'}(\momentum)}^2 \nonumber \\
&= \bigg|\frac{2ma^2}{\hbar^2} \frac{ V_{\lead} }{K_{\lead}(\momentum)} +1 \bigg|^2 + \frac{1}{\abs{K_{\lead}(\momentum)}^2}\sum^N_{\lead' \neq \lead}  \bigg(\frac{2ma^2}{\hbar^2}  V_{\lead'} \bigg)^2 \nonumber \\
&=  \bigg|\frac{2ma^2}{\hbar^2} \frac{ V_{\lead} }{\abs{K_{\lead}(\momentum)} e^{i\phi(\momentum)}} +1 \bigg|^2 + \frac{1}{\abs{K_{\lead}(\momentum)}^2} \sum^N_{\lead' \neq \lead}  \bigg(\frac{2ma^2}{\hbar^2} V_{\lead'}\bigg)^2,  \nonumber
\end{align}
where $\phi(\momentum)$ is the argument of $K_\lead(\momentum)$, and we assume $V_\lead \in \mathbb{R}$ $\forall \lead$. This gives an expression for $\abs{K_{\lead}(\momentum)}$:

\begin{align}
\abs{K_{\lead}(\momentum)} = - \frac{ma^2}{\hbar^2} \frac{  V^2_{\lead} +  \sum^N_{\lead' \neq \lead} V^2_{\lead'}}{  V_{\lead} \cos(\phi(\momentum))} =  - \frac{ma^2}{\hbar^2} \frac{1}{\cos(\phi(\momentum))}   \sum^N_{\lead'  = 1} \bigg(\frac{V_{\lead'}}{V_{\lead}}  \bigg) V_{\lead'}. \nonumber
\end{align}
Let us consider the situations where the molecule is attached to either three or four leads. First, consider $N =3$ and let $\lead = 1$ be the input lead. The total transmission for the two output leads is then

\begin{align}
\abs{T_{12}(\momentum)}^2 + \abs{T_{13}(\momentum)}^2 &=  \bigg | \frac{2ma^2}{\hbar^2}  \frac{V_{2}}{K_1(\momentum)}  \bigg |^2 +  \bigg | \frac{2ma^2}{\hbar^2}  \frac{V_{3}}{K_1(\momentum)}  \bigg |^2 \nonumber \\
& = 4 \cos^2(\phi(\momentum))\frac{(V_2/V_1)^2 + (V_3/V_1)^2}{\big( 1 + (V_2/V_1)^2 + (V_3/V_1)^2\big)^2}. \nonumber
\end{align}
To obtain high transmission, let $(V_2/V_1)^2 + (V_3/V_1)^2 = 1$, for example,  if $V_2 = V_3$, then the ratio of output to input couplings should be $V_2/V_1 = V_3/V_1=1/\sqrt{2}$, and the total transmission equals 

\begin{align}
 \abs{T_{12}(\momentum)}^2 + \abs{T_{13}(\momentum)}^2 &= \cos^2(\phi(\momentum)) && \text{if $(V_2/V_1)^2 + (V_3/V_1)^2 = 1$}. \nonumber
\end{align}
The electron would be fully transmitted, i.e., zero reflection, if $\phi(\momentum) = n \pi$ for $n = 0,\pm 1,\pm 2 \hdots$ In other words, perfect transmission is possible if the incoming momentum can be picked so that $K_\lead(\momentum)$ is real. Next, consider $N =4$. The total transmission for two output leads is then

\begin{align}
\abs{T_{13}}^2 + \abs{T_{14}}^2 &=4 \cos^2(\phi(\momentum)) \frac{(V_3/V_1)^2 + (V_4/V_1)^2 }{\big( 1 + (V_2/V_1)^2 + (V_3/V_1)^2  + (V_4/V_1)^2\big)^2}.  \label{eq:app:trans_N4}
\end{align}
This equal $\cos^{2}(\phi(\momentum))$ if and only if $ (V_2/V_1)^2 = 0$ and $V_4/V_1 = \pm \sqrt{1- (V_3/V_1)^2}$. However,  we require $ V_2 \neq 0$, otherwise the lead $\lead  = 2$ would be removed entirely from the molecule. Therefore having four leads, there will always be some reflection present.

\subsection{Electron Transmission Through Molecular Hydrogen in a Minimal Basis}
\label{app:example_h2_transmission}
\begin{figure}[H] 
\centering  
\includegraphics[width=1.0\textwidth]{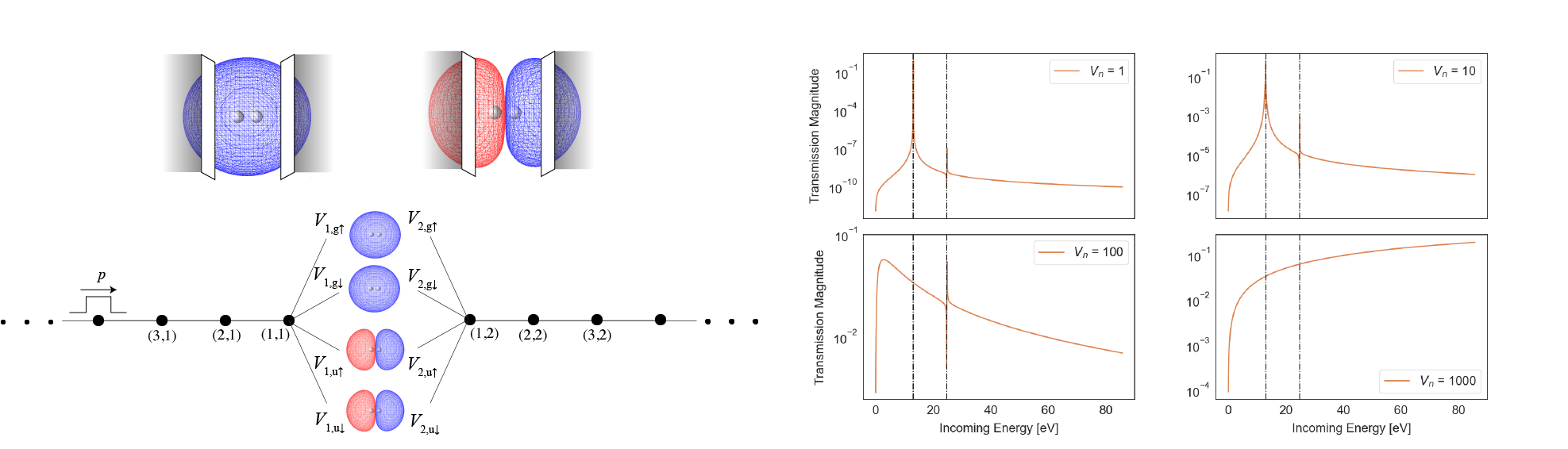}
\caption{ Electron transmission through molecular hydrogen in STO-3G basis.  Upper left: The planes intersecting through the two orbitals indicate the integration limits in Eqs. \eqref{eq:app:h2_V1} and \eqref{eq:app:h2_V2}, and the grey area indicates the volume integrated over. We choose to integrate up to $0.1a_0$ from the protons, which result in integrals of $V_{\chi_g} = 0.25$ and $V_{\chi_u} = 0.39$.  Lower left: Illustration of the molecular junction. Right: Electron transmission magnitude as a function of incoming kinetic energy ($E_{\text{in}} = \frac{\momentum^2}{2m}$) for different values of $V_{\lead}$ in  Eqs. \eqref{eq:app:h2_V1} and \eqref{eq:app:h2_V2}, and assuming $V_{\lead = 1} = V_{\lead = 2}$. We require that $\frac{a\abs{\momentum}}{\hbar} \ll 1$ (section \ref{app:kinetics}), and the quantum dot distance is set to $0.1\text{\AA}$. The vertical lines correspond to $E^{(3)}_{0} - E^{(2)}_0 $ and $E^{(3)}_{1} - E^{(2)}_0 $. }
\label{fig:H2_STO3G_transmission}
\end{figure}

In this example, we will calculate the electron transmission for molecular hydrogen in a minimal basis. For molecular hydrogen in a minimal basis, each hydrogen atom contributes one spatial orbital, and there are two possible spins for each orbital --- a total of 4 spin-orbitals (molecular orbitals). We denote these spin-orbitals as $\ket{g\uparrow, g \downarrow, u\uparrow, u\downarrow}$, where the letters \emph{g} and \emph{u} stand for gerade (even) and ungerade (odd), respectively, and the arrows denote the spin-state.  The ``exact'' energy states and energies for molecular hydrogen in the STO-3G basis at a proton-proton distance of 0.7414 $\text{\AA}$ and 1.175 $\text{\AA}$ (equilibrium bond lengths) for the uncharged and charged  molecule, respectively, are given by
 
 \begin{align}
\ket{E^{(2)}_0} &= -0.99\ket{1100} + 0.11\ket{0011} &E^{(2)}_0 = -30.91 \text{eV} \nonumber \\
\ket{E^{(3)}_0}&=\ket{1110} &E^{(3)}_0=-17.86\text{eV} \nonumber \\
\ket{E^{(3)}_1}&=\ket{1101} &E^{(3)}_1=-17.86\text{eV} \nonumber \\
\ket{E^{(3)}_2}&=\ket{1011} &E^{(3)}_2=-6.169\text{eV} \nonumber \\
\ket{E^{(3)}_3}&=\ket{0111} &E^{(3)}_3=-6.169\text{eV}. \nonumber
 \end{align}
Assuming one input and one output lead (see Fig. \ref{fig:H2_STO3G_transmission}), the coupling matrix \emph{B} and the molecular Hamiltonian \emph{D} are given by

\begin{align}
 B&=  \begin{pmatrix}
- V^*_{1,u\uparrow} \cdot 0.99 &  - V^*_{2,u\uparrow} \cdot 0.99  \\
- V^*_{1,u\downarrow} \cdot 0.99 &  - V^*_{2,u\downarrow} \cdot 0.99  \\
  V^*_{1,g\uparrow} \cdot 0.11 &    V^*_{2,g\uparrow} \cdot 0.11  \\
  V^*_{1,g\downarrow} \cdot 0.11  & V^*_{2,g\downarrow} \cdot 0.11  \\
 \end{pmatrix}, \nonumber \\
 D&=  \begin{pmatrix}
-17.86\text{eV} & 0 & 0 & 0  \\
0&  -17.86\text{eV} & 0 & 0  \\
0 & 0& -6.169\text{eV} & 0  \\
0 & 0 & 0 & -6.169\text{eV}  \\
 \end{pmatrix}.\nonumber
\end{align}
According to section \ref{app:lead_molecule}, we assume that the coupling strengths can be calculated from

\begin{align}
V_{\lead,g\sigma} =  V_{\lead} \int_{\mathcal{V}} d\vec{r} \hspace{0.1cm} \abs{\chi_{g\sigma}(\vec{r})}^2 \label{eq:app:h2_V1} \\
V_{\lead,u\sigma} = V_{\lead} \int_{\mathcal{V}} d\vec{r} \hspace{0.1cm} \abs{\chi_{u\sigma}(\vec{r})}^2. \label{eq:app:h2_V2}
\end{align}
From these matrices, we can calculate the scattering matrix \eqref{eq:app:S_matrix}. We show electron transmission as a function of incoming kinetic energy, and for different values of $V_{\lead}$, in Fig. \ref{fig:H2_STO3G_transmission}. As expected, we observe that the transmission peaks when the incoming kinetic energy match an energy eigenvalue of the charged molecule, i.e. when $\frac{\momentum^2}{2m}= E^{(3)}_{g} - E^{(2)}_0 $, as can be seen from the two plots in the upper panel. Interestingly, the peak located at the ground state shifts with increasing coupling strength, while the peak located at the excited state does not, as observed in the lower left plot. For increasing coupling strengths, the interference effects start to disappear, and we observe a more 'classical' transmission spectrum characterized by increasing transmission with increasing kinetic energy.

/
\section{Scattering of Two Electrons on Different Leads}
\label{app:S_matrix_two_particles}

To implement two-qubit gates, we will consider interactions between electrons traveling on different leads. Our starting point is the Hamiltonian  
\begin{align}
\hat{H} &=\hat{H}_{\text{leads}}  +  \hat{H}_{\text{int-leads}}, \label{eq:app:hamil_twobody_scatt}
\end{align}
where $\hat{H}_{\text{leads}}$ is the one-body Hamiltonian \eqref{eq:dis_hamil} except the leads are taken to be infinite in both ends, and we propose the following two-body term

\begin{align}
 \hat{H}_{\text{int-leads}} = \sum_{ \site_1,\site_2 \in \mathbb{Z}}V(r,d) \hat{c}_{\lead_1, \site_1 }^\dagger \hat{c}_{\lead_2, \site_2 }^\dagger \hat{c}_{\lead_2, \site_2 } \hat{c}_{\lead_1, \site_1 }, \label{eq:app:two_body}
\end{align}
with $\lead_1 \neq \lead_2$ as the lead indices, and

\begin{align}
V(r,d) =
    \begin{cases}
       K \frac{1}{\sqrt{r^2 + d^2}} & \text{if $C\geq \abs{r}$}\\
      0 & \text{if $C< \abs{r}$}
    \end{cases}        \label{eq:app:potential}
\end{align}
is the Coulomb repulsion between the two electrons. Here $r = \site_1 - \site_2 $ is the horizontal distance between the electrons (see Fig. \ref{fig:introduction_illustration}), $d\in \mathbb{Z}$ is the distance between the two leads $\lead_1$ and $\lead_2$ in units of \emph{a}, and $K = kq^2/a$ ($k=\text{Coulomb constant}$, $q=-e$, $a=\text{quantum dot distance}$). The two-body term \eqref{eq:app:two_body} assumes the electrons cannot jump to other leads, but that they can still interact while moving past each other on separate leads. The justification for this form is that the electrons would have to overcome a very large potential barrier in order to jump to other leads. Furthermore, we assume the potential \eqref{eq:app:potential} has finite range \emph{C}, which means that $V(r,d) = 0$ whenever $C< \abs{r}$ (same assumption as in Ref. \cite{childs_universal_2013}, SI section S2).  The outline of this section is as follows: 

\begin{enumerate}
    \item Transform the Hamiltonian \eqref{eq:app:hamil_twobody_scatt} into an effective one-body Hamiltonian [Eq. \eqref{eq:app:effec:hamil}]. 
    \item Propose an effective one-particle scattering ansatz [Eq. \eqref{eq:app:effec_scat_state}]
    \item Solve for the scattering coefficients [Eqs. \eqref{eq:app:firsteq}-\eqref{eq:app:lasteq}].
\end{enumerate}
We first transform the Hamiltonian into a different basis wherein we describe the space $(\site_1,\site_2) \in \mathbb{Z}^2$ using the following coordinates 

\begin{align}
s(\site_1, \site_2) &= \site_1 + \site
_2 \nonumber \\
r(\site_1, \site_2) &= \site_1 - \site
_2 \nonumber \\
\mathcal{S} &= \text{$\{(s,r) \in \mathbb{Z}^2:$ \emph{s} and \emph{r} are both even or both odd\}}. \nonumber 
\end{align}
This basis transformation was proposed in Ref. \cite{childs_universal_2013} (SI, section S2) to evaluate the scattering of two electrons on the same lead. This transformation is an injective map from $(\site_1,\site_2) \in \mathbb{Z}^2$ to $(s,r) \in \mathcal{S}$, meaning each set of values of $\site_i,\site_2$ produce an unique pair of values for \emph{s} and \emph{r}. For instance, this means we can think of the state $\ket{\site_1 = 5, \site_2 = 2}$ as being simply renamed to $\ket{s=7, r=3}$. Thus assuming we have one particle on each of the leads $\lead_1, \lead_2$, we can rewrite the operators as follows

\begin{align}
\hat{H}^{(\site_1)}_{\text{leads}} \otimes \mathbb{1}^{(\site_2)} =& \sum_{\site_1,\site_2 \in \mathbb{Z}} \bigg(  2 \beta \hat{c}^\dagger_{\lead_1,\site_1} \hat{c}_{\lead_1,\site_1} \otimes  \hat{c}^\dagger_{\lead_2,\site_2} \hat{c}_{\lead_2,\site_2} \nonumber \\
&-\beta \big(\hat{c}^\dagger_{\lead_1,\site_1} \hat{c}_{\lead_1,\site_1 + 1} \otimes  \hat{c}^\dagger_{\lead_2,\site_2} \hat{c}_{\lead_2,\site_2} + \hat{c}^\dagger_{\lead_1,\site_1+1} \hat{c}_{\lead_1,\site_1} \otimes  \hat{c}^\dagger_{\lead_2,\site_2} \hat{c}_{\lead_2,\site_2} \big) \bigg) \nonumber \\
=& \sum_{(s,r) \in \mathcal{S}}  \bigg(  2 \beta \hat{c}^\dagger_{\lead_1,\lead_2,s} \hat{c}_{\lead_1,\lead_2,s} \otimes  \hat{c}^\dagger_{\lead_1,\lead_2,r} \hat{c}_{\lead_1,\lead_2,r} \nonumber \\
&-\beta \big(\hat{c}^\dagger_{\lead_1,\lead_2, s} \hat{c}_{\lead_1,\lead_2, s + 1} \otimes  \hat{c}^\dagger_{\lead_1,\lead_2, r} \hat{c}_{\lead_1,\lead_2, r + 1} +\hat{c}^\dagger_{\lead_1,\lead_2, s+1} \hat{c}_{\lead_1,\lead_2, s} \otimes  \hat{c}^\dagger_{\lead_1,\lead_2, r + 1}  \hat{c}_{\lead_1,\lead_2, r}\big) \bigg). \nonumber 
\end{align}
A similar expression exists for $ \mathbb{1}^{(\site_1)} \otimes \hat{H}^{(\site_2)}_{\text{leads}}$. The map is in fact a bijection, meaning there is a 1:1 correspondence between each pair $(s,r) \in \mathcal{S}$ and each pair of values $(\site_1,\site_2) \in \mathbb{Z}^2$. In other words, summing over all values of  $(\site_1,\site_2) \in \mathbb{Z}^2$ yields exactly the same states as summing over all of the values of $(s,r) \in \mathcal{S}$. The interaction term similarly becomes

\begin{align}
 \hat{H}_{\text{int-leads}} = \sum_{ (s,r) \in \mathcal{S}}V(r,d) \hat{c}_{\lead_1, \lead_2,s }^\dagger \hat{c}_{\lead_1, \lead_2,r }^\dagger \hat{c}_{\lead_1, \lead_2,r } \hat{c}_{\lead_1, \lead_2,s }. \nonumber
\end{align}
The definition of the coordinates means only the pairs of \emph{s} and \emph{r} within the set $\mathcal{S}$ actually correspond to valid positions $\site_1$ and $\site_2$. If other integer values of \emph{s} and \emph{r} are used, the inverse mapping back to $\site_1$ and $\site_2$-coordinates will yield invalid half-integer values of  $\site_1$ and $\site_2$, e.g. $\site_1 = \frac{3}{2}$ and $\site_2 = \frac{1}{2}$. In other words, if we just sum the indexes $(s,r)$ over the entirety of $\mathbb{Z}^2$ in the equations above, we will accidentally include a lot of sets of indexes that do not correspond to valid states. However, let us assume we artificially extend the Hilbert space to also include these invalid states, i.e., we now have the following two disjoint classes of states:

\begin{align}
&\{\ket{s,r}: (s,r)\in \mathcal{S}\} &      \text{valid states} \nonumber \\
&\{\ket{s,r}: (s,r)\in \mathbb{Z}^2 \setminus \mathcal{S}\}  &      \text{invalid states.} \nonumber 
\end{align}
With this distinction in mind, consider now an operator of the form $c^\dagger_{s+1} c_{s} \otimes  c^\dagger_{r + 1}  c_{r}$ for some arbitrary integer values of \emph{s} and \emph{r}. It can never introduce any mixing between the two types of states, and adding operators of this form with invalid pairs of \emph{s} and \emph{r} will thus only induce mappings among the invalid states, leaving the valid states unchanged. If we promise not to apply the Hamiltonian to invalid states, we may therefore just as well use $(s,r)\in \mathbb{Z}^2$. We thus arrive at the following expression for the Hamiltonian

\begin{align}
\hat{H}&=  4 \beta  \sum_{s,r \in \mathbb{Z}}  \bigg(  \hat{c}^\dagger_{\lead_1,\lead_2,s} \hat{c}_{\lead_1,\lead_2,s} \otimes  \hat{c}^\dagger_{\lead_1,\lead_2,r} \hat{c}_{\lead_1,\lead_2,r}\bigg) \nonumber \\
&-\beta\sum_{s,r \in \mathbb{Z}}  \bigg(\hat{c}^\dagger_{\lead_1,\lead_2, s} \hat{c}_{\lead_1,\lead_2, s + 1} \otimes  \hat{c}^\dagger_{\lead_1,\lead_2, r} \hat{c}_{\lead_1,\lead_2, r + 1} +
\hat{c}^\dagger_{\lead_1,\lead_2, s+1} \hat{c}_{\lead_1,\lead_2, s} \otimes  \hat{c}^\dagger_{\lead_1,\lead_2, r + 1}  \hat{c}_{\lead_1,\lead_2, r} \bigg) \nonumber \\
&-\beta  \sum_{s,r \in \mathbb{Z}}  \bigg(\hat{c}^\dagger_{\lead_1,\lead_2, s} \hat{c}_{\lead_1,\lead_2, s + 1} \otimes  \hat{c}^\dagger_{\lead_1,\lead_2, r} \hat{c}_{\lead_1,\lead_2, r - 1} +
\hat{c}^\dagger_{\lead_1,\lead_2, s+1} \hat{c}_{\lead_1,\lead_2, s} \otimes  \hat{c}^\dagger_{\lead_1,\lead_2, r - 1}  \hat{c}_{\lead_1,\lead_2, r} \bigg) \nonumber \\
&+\sum_{ (s,r) \in \mathbb{Z}}V(r,d) \hat{c}_{\lead_1, \lead_2,s }^\dagger \hat{c}_{\lead_1, \lead_2,r }^\dagger \hat{c}_{\lead_1, \lead_2,r } \hat{c}_{\lead_1, \lead_2,s }. \label{eq:app:transform}
\end{align}
Shifting the index of the \emph{r}-summation on the third sum in Eq. \eqref{eq:app:transform} and looking closely at the contributions from the hopping terms, we see that the Hamiltonian can be rewritten as the following expression:

\begin{align}
\hat{H} &=  4\beta \mathbb{1}_s \otimes \mathbb{1}_r   \nonumber \\
&-\beta \sum_{s \in \mathbb{Z}} \big[ \hat{c}^\dagger_{\lead_1,\lead_2, s} \hat{c}_{\lead_1,\lead_2, s + 1} + \hat{c}^\dagger_{\lead_1,\lead_2, s + 1}  \hat{c}_{\lead_1,\lead_2, s} \big] \otimes  \sum_{r \in \mathbb{Z}} \big[ \hat{c}^\dagger_{\lead_1,\lead_2, r} \hat{c}_{\lead_1,\lead_2, r + 1} + \hat{c}^\dagger_{\lead_1,\lead_2, r + 1}  \hat{c}_{\lead_1,\lead_2, r} \big] \nonumber \\
&+\mathbb{1}_s \otimes  \sum_{ r \in \mathbb{Z}}V(r,d)  \hat{c}_{\lead_1, \lead_2,r }^\dagger \hat{c}_{\lead_1, \lead_2,r }. \label{eq:app:effec:hamil}
\end{align}
The transformed Hamiltonian has the nice property that it does not mix \emph{s} and \emph{r}. The \emph{s}-part is simply an effective one-body Hamiltonian describing an infinite chain with nearest neighbour interaction, which has known solutions. Specifically, the state 

\begin{align}
\sum_{s\in \mathbb{Z}} e^{-i \frac{\momentum_1 a s }{2\hbar}} \ket{\lead_1,\lead_2,s}
\end{align}
is an eigenstate of the only non-trivial operator in \emph{s}-space. This leaves the \emph{r}-part, which includes the two-body interaction. For each $\momentum_1 \in \mathbb{R}$ and $\momentum_2\leq 0$, we define a scattering state of the form 

\begin{figure}[t] 
\centering  
\includegraphics[width=0.6\textwidth]{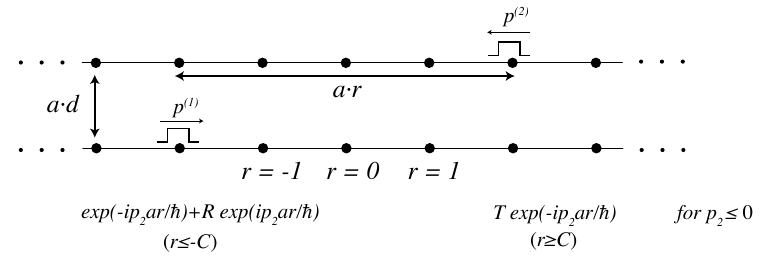}
\caption{ Two-particle scattering. The momentum in \emph{r}-space is $\momentum_2 = (\momentum^{(1)} - \momentum^{(2)})/2$, where $\momentum^{(1)}$ and  $\momentum^{(2)}$ are the individual momenta of the electrons.  }
\label{fig:app:two_particles_scattering_illu}
\end{figure}

\begin{align}
\braket{s;r,\lead'_2,\lead'_1|\text{sc}_{\lead_1, \lead_2}(\momentum_1;\momentum_2)}  = e^{-i \frac{\momentum_1 a s }{2\hbar}} \delta_{\lead'_1,\lead_1}\delta_{\lead'_2,\lead_2}
    \begin{cases}
        e^{-i \frac{\momentum_2 a r}{\hbar}}  + R(\momentum_1 ,\momentum_2)  e^{i \frac{\momentum_2 a r}{\hbar}}& r \leq -C \\
      f(\momentum_1 ,\momentum_2, r) & \abs{r} < C\\
     T(\momentum_1 ,\momentum_2)  e^{-i \frac{\momentum_2 a r}{\hbar}} & r \geq C. 
    \end{cases} \label{eq:app:effec_scat_state}
\end{align}
Antisymmetrization of the wave function is not necessary, which we explain in greater detail in section \ref{sec:app:antisymmetrization}. Why the factor 1/2 in the exponent for the \emph{s}-part? The two indexes \emph{s} and \emph{r} in this system are restricted to only take values where $s = \site_1 + \site_2$ and $r = \site_1 - \site_2$. This means we can pick one index freely, and then just implement the $\mathcal{S}$-set restriction on the other index.  If we allow \emph{r} to be arbitrary, the smallest translation we can do on \emph{s} is by two units, $\hat{T}_{s\rightarrow s+2}\ket{s;r} = \ket{s+2;r}$, otherwise we move out of the allowed values of \emph{s} and \emph{r}. Thus the factor 1/2 makes the phase of the smallest possible translation to be $\exp(\pm i \momentum_1 a /\hbar)$. Similarly, the factors $\exp(\pm i \frac{\momentum_2 a r}{\hbar})$ are picked up by the wavefunction in \eqref{eq:app:effec_scat_state} when applying the discretized translation operator in \emph{r}-space to move the particles. 

Let $\ket{\psi_{\lead_1,\lead_2}(\momentum_1;\momentum_2})$ consist of the \emph{r}-part of the scattering wave function in \eqref{eq:app:effec_scat_state}. Firstly, consider the hopping term in \eqref{eq:app:effec:hamil} for the \emph{r}-part without the constant factor $-\beta$: 

\begin{align}
& \sum_{r \in \mathbb{Z}} \big[ \hat{c}^\dagger_{\lead'_1,\lead'_2, r} \hat{c}_{\lead'_1,\lead'_2, r + 1} + \hat{c}^\dagger_{\lead'_1,\lead'_2, r + 1}  \hat{c}_{\lead'_1,\lead'_2, r} \big] \ket{\psi_{\lead_1,\lead_2}(\momentum_1 ;\momentum_2}) \nonumber \\
&=  \sum_{r \leq -C}  \big(e^{-i \frac{\momentum_2 a r}{\hbar}}  + R e^{i \frac{\momentum_2 a r}{\hbar}}\big) \big(\ket{\lead_1,\lead_2,r+1} + \ket{\lead_1,\lead_2,r-1} \big) \nonumber \\
&+ \sum_{r \geq C} T e^{-i \frac{\momentum_2 a r}{\hbar}} \big(\ket{\lead_1,\lead_2,r+1} + \ket{\lead_1,\lead_2,r-1} \big) \nonumber \\
&+\sum_{\abs{r} < C} f(r) \big(\ket{\lead_1,\lead_2,r+1} + \ket{\lead_1,\lead_2,r-1} \big) \nonumber \\
&= 2\cos\bigg(\frac{\momentum_2 a}{\hbar}\bigg) \sum_{r \leq -C}  \big(e^{-i \frac{\momentum_2 a r}{\hbar}}  + R e^{i \frac{\momentum_2 a r}{\hbar}}\big) \ket{r} +  \big(e^{i \frac{\momentum_2 a C}{\hbar}}  + R e^{-i \frac{\momentum_2 a C}{\hbar}}\big) \ket{-C+1} \nonumber \\
&- \big(e^{-i \frac{\momentum_2 a (-C+1)}{\hbar}}  + R e^{i \frac{\momentum_2 a (-C+1)}{\hbar}}\big) \ket{-C} +2\cos\bigg(\frac{\momentum_2 a}{\hbar}\bigg) \sum_{r \geq C}   T e^{-i \frac{\momentum_2 a r}{\hbar}}\ket{r} -  T e^{-i \frac{\momentum_2 a (C-1)}{\hbar}}\ket{C} \nonumber \\
&+ T e^{-i \frac{\momentum_2 a C}{\hbar}}\ket{C-1} + \sum_{\abs{r} < C} \bigg( f(r-1) +  f(r+1)\bigg) \ket{r} - f(-C) \ket{-C+1} + f(C-1) \ket{C} \nonumber \\
&+  f(-C+1) \ket{-C} - f(C) \ket{C-1}, \nonumber 
\end{align}
where we omitted the lead indices, i.e. $\ket{\lead_1,\lead_2,r} \rightarrow \ket{r}$. Similarly, the interaction part gives 

\begin{align}
 &\sum_{ r \in \mathbb{Z}}V(r,d)  \hat{c}_{\lead'_1, \lead'_2,r }^\dagger \hat{c}_{\lead'_1, \lead'_2,r }\ket{\psi_{\lead_1,\lead_2}(\momentum_1 ,\momentum_2}) \nonumber \\
 &=\sum_{\abs{r} < C} f(r) V(r,d) \ket{r} + V(-C,d) \big(e^{i \frac{\momentum_2 a C}{\hbar}}  + R e^{-i \frac{\momentum_2 a C}{\hbar}}\big) \ket{-C}+ V(C,d) T e^{-i \frac{\momentum_2 a C}{\hbar}} \ket{C}. \nonumber 
\end{align}
Based on the above equations, we require that 

\begin{align}
\hat{H} \ket{\text{sc}_{\lead_1, \lead_2}(\momentum_1;\momentum_2)} &= 4\beta\bigg(1 -  \cos(\frac{\momentum_1 a}{2\hbar})\cos(\frac{\momentum_2 a}{\hbar}) \bigg)\ket{\text{sc}_{\lead_1, \lead_2}(\momentum_1;\momentum_2)}, \nonumber 
\end{align}
which is true if and only if we can pick the parameters $\{R,T,f\}$ such that:

\begin{align}
 & -2\beta\cos\bigg(\frac{\momentum_1 a}{2\hbar}\bigg)\bigg(f(C-1) -T e^{-i \frac{\momentum_2 a (C-1)}{\hbar}}\bigg) + V(C,d) T e^{-i \frac{\momentum_2 a C}{\hbar}} =  0 \label{eq:app:firsteq} \\[0.3cm]
&- 2\beta\cos\bigg(\frac{\momentum_1 a}{2\hbar}\bigg) \bigg( T e^{-i \frac{\momentum_2 a C}{\hbar}} +f(C-2) \bigg) + f(C-1) V(C-1,d) \nonumber \\
&= -4\beta \cos\bigg(\frac{\momentum_1 a}{2\hbar}\bigg) \cos\bigg(\frac{\momentum_2 a}{\hbar}\bigg) f(C-1)  \\[0.3cm] 
&- 2\beta\cos\bigg(\frac{\momentum_1 a}{2\hbar}\bigg) \bigg(f(C-3) + f(C-1)\bigg) + f(C-2)V(C-2,d) \nonumber \\
&= -4 \beta \cos\bigg(\frac{\momentum_1 a}{2\hbar}\bigg) \cos\bigg(\frac{\momentum_2 a}{\hbar}\bigg) f(C-2)  \\[0.3cm]
 &\vdots \nonumber \\[0.3cm]
&  -2\beta\cos\bigg(\frac{\momentum_1 a}{2\hbar}\bigg) \bigg(f(-C+1) + f(-C+3)\bigg)   + f(-C+2)V(-C+2,d) \nonumber \\
&= -4 \beta \cos\bigg(\frac{\momentum_1 a}{2\hbar}\bigg) \cos\bigg(\frac{\momentum_2 a}{\hbar}\bigg) f(-C+2) \\[0.3cm] 
&-2\beta\cos\bigg(\frac{\momentum_1 a}{2\hbar}\bigg) \bigg(  e^{i \frac{\momentum_2 a C}{\hbar}} +  Re^{-i \frac{\momentum_2 a C}{\hbar}} + f(-C+2) \bigg) + f(-C+1) V(-C+1,d) \nonumber \\
&=-4 \beta \cos\bigg(\frac{\momentum_1 a}{2\hbar}\bigg) \cos\bigg(\frac{\momentum_2 a}{\hbar}\bigg) f(-C+1) \\[0.3cm] 
&-2\beta \cos\bigg(\frac{\momentum_1 a}{2\hbar}\bigg)\bigg(  f(-C+1) - \bigg(  e^{-i \frac{\momentum_2 a (-C+1)}{\hbar}} +  Re^{i \frac{\momentum_2 a (-C+1)}{\hbar}}\bigg)  \bigg)   \nonumber \\
&+  V(-C,d) \big(e^{i \frac{\momentum_2 a C}{\hbar}}  + R e^{-i \frac{\momentum_2 a C}{\hbar}}\big) = 0. \label{eq:app:lasteq}
\end{align}
There are a total of $2C+1$ equations and the same number of unknowns. Note that we can express the momentum in $s,r$-space in terms of momentum in site-space by $\momentum_1 = \momentum^{(1)} + \momentum^{(2)}$ and $\momentum_2 = (\momentum^{(1)} - \momentum^{(2)})/2$, where $\momentum^{(1)}$ and $\momentum^{(2)}$ are the individual momenta.  We can assume the two electrons share the same momentum. In that case, $\momentum_1 = 0$ and $\momentum_2 =  \momentum^{(1)}$, since  $ \momentum^{(1)} = - \momentum^{(2)} $ (one electron propagates in opposite direction). We require that $\beta = \hbar^2/2ma^2$ and $a\momentum^{(1)},a\momentum_2\ll \hbar$ (or $a \momentum_2 / \hbar\ll 1$) (section \ref{app:kinetics}). The transmission coefficient is obtained by solving Eqs. \eqref{eq:app:firsteq}-\eqref{eq:app:lasteq}.

\begin{figure}[t] 
\centering  
\includegraphics[width=0.9\textwidth]{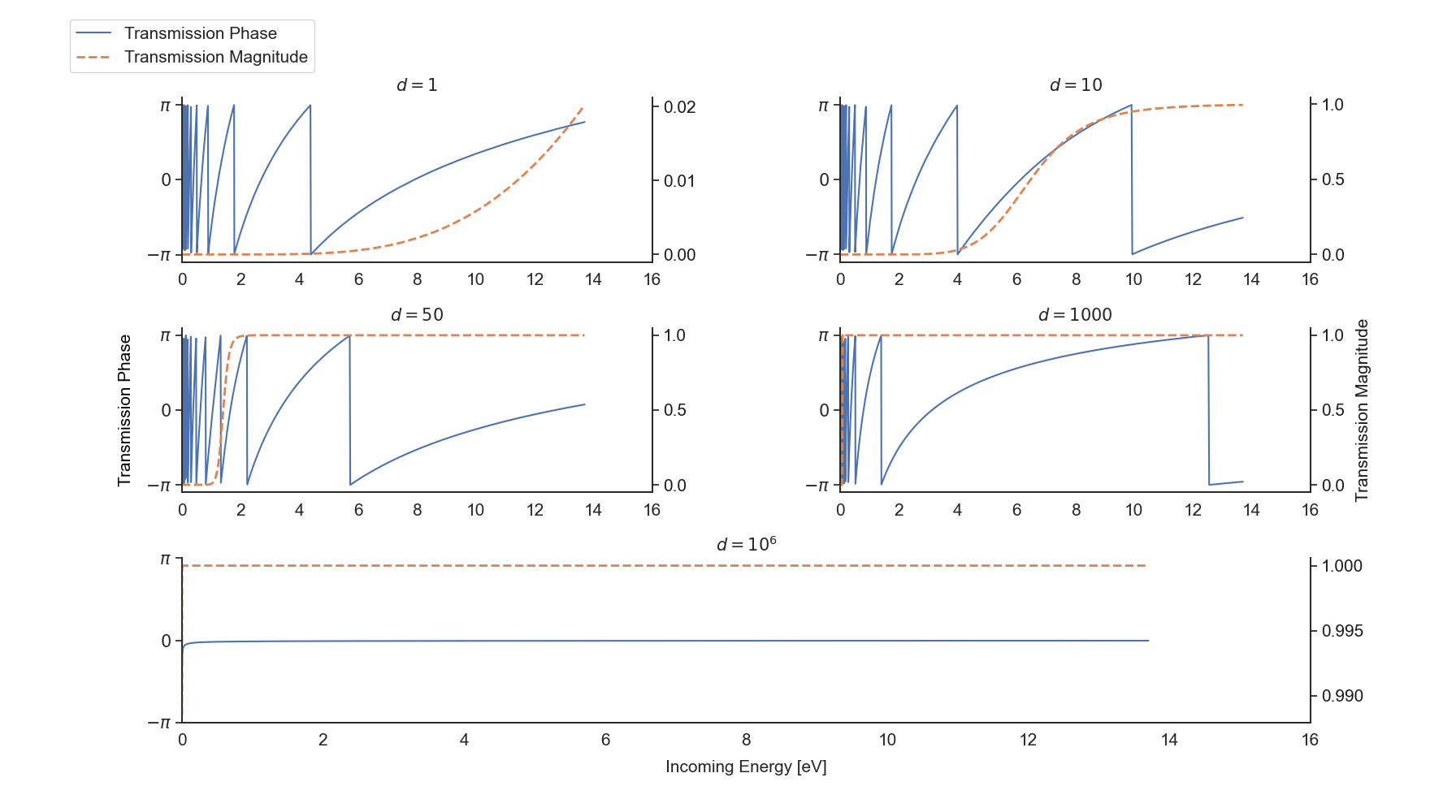}
\caption{ Two-electron transmission magnitude $\abs{T}^2$ and phase $\arg(T)$ for the situation  depicted in Fig. \ref{fig:app:two_particles_scattering_illu}, as a function of incoming kinetic energy, and for  different values of the distance between the leads. The electrons are assumed to have the same kinetic energy, and for the potential \eqref{eq:app:potential} we set $C = 10^4$.  }
\label{fig:app:two_particles_scattering_numerics}
\end{figure}

Figure \ref{fig:app:two_particles_scattering_numerics} shows two-electron transmission magnitude and phase when letting the electrons past each other on separate leads while interacting for different values of the distance between the leads. The case where $d = 1000$ is also shown in the main text in Fig. \ref{fig:introduction_illustration}. If the lead-lead distance equals $d = 10^6$, i.e., $10^6\cdot a = 10^4$nm where $a = 0.1\text{\AA}$, the interaction between the electrons is very weak, and the transmission magnitude is basically a step function and the transmission phase becomes zero. The reason that $\arg(T)\rightarrow 0$ for $d \rightarrow \infty$ is explained in the following section \ref{app:example_lead_effective_length_r_space}.

\subsection{Transmission Phase Through a Linear Chain of Quantum Dots in (\emph{r},\emph{s})-Space}
\label{app:example_lead_effective_length_r_space}
\begin{figure}[H] 
\centering  
\includegraphics[width=0.6\textwidth]{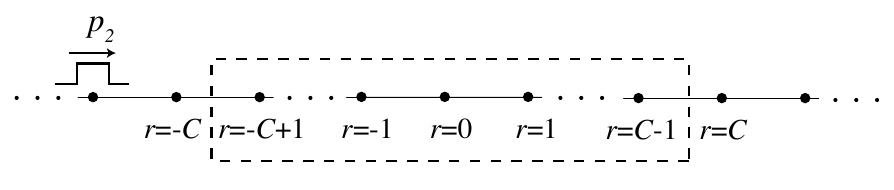}
\caption{ Schematic depiction of a linear chain of dots in \emph{r}-space with each dot representing a quantum state coupled to its two nearest neighbours. The momentum in \emph{r}-space is $\momentum_2 = (\momentum^{(1)} - \momentum^{(2)})/2$, where $\momentum^{(1)}$ and  $\momentum^{(2)}$ are the individual momenta of the electrons.}
\label{fig:linear_chain_r_space}
\end{figure}

In this example, we will find an expression for the transmission amplitude and phase through a straight lead in ($r,s$)-space. Assume the distance between the two leads is very large, $d\gg 1$, such that $V(C,d) = 0$. In this case, the \emph{r}-part of the Hamiltonian \eqref{eq:app:effec:hamil} is simply an effective one-body Hamiltonian describing an infinite chain with nearest neighbor interaction, as depicted in Fig. \ref{fig:linear_chain_r_space}. Furthermore, we assume the electrons have the same momentum and going in opposite direction, such that $\momentum_1 = 0$. Equations \eqref{eq:app:firsteq}-\eqref{eq:app:lasteq} are then simplified to

\begin{align}
&f(C-1) - T e^{-i \frac{\momentum_2 a}{\hbar} (C-1)}  = 0 \label{app:eq:linear_chain_r_space_eq1} \\
&T e^{-i \frac{\momentum_2 a}{\hbar} C} + f(C-2) = \bigg(  e^{ \frac{\momentum_2 a}{\hbar}}  + e^{ -\frac{\momentum_2 a}{\hbar}} \bigg) f(C-1) \label{app:eq:linear_chain_r_space_eq2} \\
&e^{i \frac{\momentum_2 a}{\hbar} C} + Re^{-i \frac{\momentum_2 a}{\hbar} C} + f(-C+2) = \bigg(  e^{ \frac{\momentum_2 a}{\hbar}}  + e^{ -\frac{\momentum_2 a}{\hbar}} \bigg) f(-C+1)  \label{app:eq:linear_chain_r_space_eq3} \\
&f(-C+1) - e^{-i \frac{\momentum_2 a}{\hbar} (-C+1)} - R  e^{i \frac{\momentum_2 a}{\hbar} (-C+1)} = 0 \label{app:eq:linear_chain_r_space_eq4}  \\
&f(r-1) + f(r+1) =  \bigg(  e^{ \frac{\momentum_2 a}{\hbar}}  + e^{ -\frac{\momentum_2 a}{\hbar}} \bigg) f(r) \quad \forall r \in \{ -C+2,\hdots,C-2\}. \label{app:eq:linear_chain_r_space_eq5} 
\end{align}
Firstly, we solve for the internal amplitudes as a function of the scattering coefficients.  A reasonable guess for a solution is the plane-wave structure of the wave function does not change when you move across the internal graph:

\begin{align}
f(r) = e^{-i \frac{\momentum_2 a}{\hbar} r} + Re^{i \frac{\momentum_2 a}{\hbar} r} && \forall r\in \{-C+1,\hdots,C-1\}. \label{app:eq:linear_chain_r_space_f} 
\end{align}
Equation \eqref{app:eq:linear_chain_r_space_f} fulfills Eqs. \eqref{app:eq:linear_chain_r_space_eq3}-\eqref{app:eq:linear_chain_r_space_eq5}, and Eqs. \eqref{app:eq:linear_chain_r_space_eq1} and \eqref{app:eq:linear_chain_r_space_eq2} if

\begin{align}
R = 0, \quad T = 1.     \nonumber 
\end{align}
Thus we have perfect transmission in the process, as expected. Notice the transmission phase is zero, whereas in section \ref{app:example_lead_effective_length} the phase equals $e^{i\frac{\momentum a (G-1)}{a}}$. The reason is how we choose to label the quantum dots in the linear chain in Fig. \ref{fig:app:two_particles_scattering_numerics} and Fig. \ref{fig:linear_chain}.

\section{Why Antisymmetrization is not Necessary}
\label{sec:app:antisymmetrization}

To construct fermionic scattering states, we antisymmetrize as follows:

\begin{align}
\ket{\text{sc}^A_{\lead_1, \lead_2}(\momentum_1; \momentum_2)} = \frac{1}{\sqrt{2}} \bigg( \ket{\text{sc}_{\lead_1, \lead_2}(\momentum_1;\momentum_2)} - \ket{\text{sc}_{\lead_2, \lead_1}(\momentum_1; -\momentum_2)} \bigg), \nonumber
\end{align}
where the first term on the right is given in Eq. \eqref{eq:app:effec_scat_state}, and for the second term the electrons are exchanged such that $\momentum_1 = \momentum^{(1)} +  \momentum^{(2)}  \rightarrow \momentum_1$ and $\momentum_2 =  (\momentum^{(1)} -  \momentum^{(2)})/2 \rightarrow - \momentum_2$.
Then

\begin{align}
&\braket{\text{sc}^A_{\lead_1, \lead_2}(\momentum_1;\momentum_2)| \hat{H}|\text{sc}^A_{\lead_1, \lead_2}(\momentum_1;\momentum_2)}  \nonumber \\
&= \frac{1}{2} \bigg( \braket{\text{sc}_{\lead_1, \lead_2}(\momentum_1;\momentum_2)| \hat{H}|\text{sc}_{\lead_1, \lead_2}(\momentum_1;\momentum_2)}+ \braket{\text{sc}_{\lead_2, \lead_1}(\momentum_1;-\momentum_2)| \hat{H}|\text{sc}_{\lead_2, \lead_1}(\momentum_1;-\momentum_2)} \nonumber \\
 &- \braket{\text{sc}_{\lead_1, \lead_2}(\momentum_1;\momentum_2)| \hat{H}|\text{sc}_{\lead_2, \lead_1}(\momentum_1;-\momentum_2)} - \braket{\text{sc}_{\lead_2, \lead_1}(\momentum_1;-\momentum_2)| \hat{H}|\text{sc}_{\lead_1, \lead_2}(\momentum_1;\momentum_2)} \bigg) \nonumber \\
 & = \braket{\text{sc}_{\lead_1, \lead_2}(\momentum_1;\momentum_2)| \hat{H}|\text{sc}_{\lead_1, \lead_2}(\momentum_1;\momentum_2)}.\nonumber
\end{align}
since the Hamiltonian \eqref{eq:app:two_body} does not contain terms which move electrons from one lead to another. Similar results also holds for all other observable relevant to the framework proposed here. This is different from the original work by Andrew Childs and co-authors in Ref. \cite{childs_universal_2013}, where they considered interaction between two electrons on the same lead. In that case, antisymmetrization does matter.

\section{Derivation of the Final Transmission Magnitude of the Molecular Circuit}
\label{app:sec:transmission_probability_of_molecular_circuit}

In the following, we will derive the final transmission magnitude given in Eq. \eqref{eq:final_transmission}. After scattering of the second molecule, the system state is 

\begin{align}
&\frac{e^{i \theta }}{\sqrt{2}}\bigg[ \bigg( T_{11} \ket{\lead = 1} + T_{12} \ket{\lead = 2} + T_{13} \ket{\lead = 3}\bigg) \nonumber \\
&+  e^{i \phi} \bigg( T_{22} \ket{\lead = 2} + T_{21} \ket{\lead = 1} + T_{23} \ket{\lead = 3}\bigg) \bigg], \nonumber
\end{align}
where we have denoted the electron being along the chain $\lead$ by the ket $\ket{\lead}$, and $\lead = 3$ is the output lead. The two input leads have same coupling strengths and therefore, due to spatial symmetry, we have

\begin{align}
T_{13} &= T_{23}    \nonumber \\
T_{12} &= T_{21}    \nonumber \\
T_{11} &= T_{22},    \nonumber 
\end{align}
and the equation reads 

\begin{align}
\frac{e^{i \theta}}{\sqrt{2}} \bigg[ \bigg( T_{11} +e^{i \phi} T_{12} \bigg)\ket{\lead = 1} + \bigg( T_{12} +e^{i \phi} T_{11} \bigg)\ket{\lead = 2}+  T_{13} \bigg( 1 +e^{i \phi}  \bigg)\ket{\lead = 3}\bigg]. \nonumber
\end{align}
The probability for finding the electron along the lead $\lead = 3$ is 

\begin{align}
\abs{T_f}^2= \abs{T_{13}}^2 \bigg( 1 + \cos( \phi) \bigg). \nonumber
\end{align}
If we assume the \emph{S}-matrix is invariant under time-reversal, then 

\begin{align}
    T_{13} = T_{31} =  \frac{e^{i\xi}}{\sqrt{2}}, \nonumber
\end{align}
and we obtain the expression in Eq. \eqref{eq:final_transmission}.
\section{Table of Assumptions}
\label{sec:app:List_of_Approximations}
\renewcommand{\arraystretch}{1.5}

\begin{table}[H]
\begin{center}
  \begin{tabular}{m{10cm}  |  m{4cm}  }
    \hline 
     The initial molecular state before scattering is the ground state.  If the experiment is performed under room temperature, the assumption is expected to be valid.\textsuperscript{\emph{a}}   & See discussion in section \ref{sec:conclusion_and_outlook}  \\ \hline
    The theory is developed within the pure state formalism.  & See discussion in section \ref{sec:conclusion_and_outlook}\\ \hline
      Molecular vibrations are not included in the theory.\textsuperscript{\emph{a}}  & See discussion in section \ref{sec:conclusion_and_outlook}   \\ \hline 
     The metallic leads are described using a free particle model. The metallic leads do not play an essential role in our description, and could be replaced by other systems. & Supporting information \ref{app:kinetics}   \\ \hline 
        Separation of indices of the coupling elements:  $V_{\lead,p} = V_{\lead} V_{p}$.\textsuperscript{\emph{a}} & Supporting information   \ref{app:lead_molecule}   \\\hline
      The coupling $V_{p}$ equals the absolute  square of the molecular orbital which  takes part of the transfer process, integrated over a region close to the leads.\textsuperscript{\emph{a}}   & Supporting information \ref{app:lead_molecule}  \\\hline
     Electrons initially bound to the molecule do not exit the molecule. & Supporting information \ref{app:One_Particle_Scattering_Matrix}, just below Eq. \eqref{app:eq:eign}\\ \hline
    The initial molecular state is not changed by the scattering (i.e. elastic scattering).\textsuperscript{\emph{a}}   & Supporting information \ref{app:One_Particle_Scattering_Matrix}, just below Eq. \eqref{app:eq:sca5} \\ \hline
  The electrons in the circuit have the same momentum. & Supporting information \ref{app:S_matrix_two_particles}  \\ \hline
    \hline
  \end{tabular}
\\\textsuperscript{\emph{a}} The theory proposed here is not limited to this description. \\
\end{center}
\caption{ Here we summarize the assumptions throughout this work.  }
\end{table}

\setkeys{acs}{maxauthors = 0}
\bibliography{references}

\end{document}